\documentstyle[epsf]{article}

\def\ra{\rightarrow}
\def\longra{\longrightarrow}

\def\fl{\forall}
\def\ify{\infty}

\def\ot{\otimes}

\def\ts{\times}
\def\wt{\widetilde}

\def\b{\beta}
\def\d{\delta}
\def\g{\gamma}

\def\t{\theta}
\def\ve{\varepsilon}
\def\vp{\varphi}

\def\G{\Gamma}

\font\tenbb=msbm10 \font\sevenbb=msbm7 \font\fivebb=msbm5
\newfam\bbfam
\textfont\bbfam=\tenbb \scriptfont\bbfam=\sevenbb
\scriptscriptfont\bbfam=\fivebb
\def\bb{\fam\bbfam}

\def\Cb{{\bb C}}

\def\Rb{{\bb R}}
\def\Qb{{\bb Q}}

\def\Kb{{\bb K}}

\catcode`\@=11
\def\displaylinesno #1{\displ@y\halign{
\hbox to\displaywidth{$\@lign\hfil\displaystyle##\hfil$}&
\llap{$##$}\crcr#1\crcr}}

\def\ldisplaylinesno #1{\displ@y\halign{
\hbox to\displaywidth{$\@lign\hfil\displaystyle##\hfil$}&
\kern-\displaywidth\rlap{$##$} \tabskip\displaywidth\crcr#1\crcr}}
\catcode`\@=12

\def\semi{\mathop{>\!\!\!\triangleleft}}

\def\build#1_#2^#3{\mathrel{
\mathop{\kern 0pt#1}\limits_{#2}^{#3}}}

\def\ot{\otimes}

\def\v{\;\raisebox{-1.5mm}{\epsfysize=6mm\epsfbox{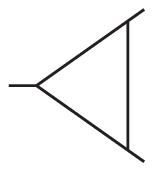}}\;}

\def\vlv{\;\raisebox{-1.5mm}{\epsfysize=6mm\epsfbox{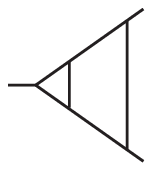}}\;}

\def\vuv{\;\raisebox{-1.5mm}{\epsfysize=6mm\epsfbox{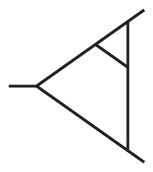}}\;}

\def\vdv{\;\raisebox{-1.5mm}{\epsfysize=6mm\epsfbox{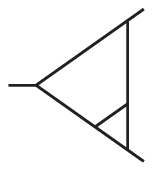}}\;}

\def\vup{\;\raisebox{-1.5mm}{\epsfysize=6mm\epsfbox{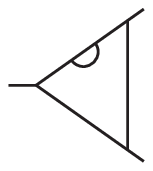}}\;}

\def\vdp{\;\raisebox{-1.5mm}{\epsfysize=6mm\epsfbox{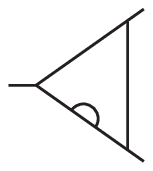}}\;}

\def\vrp{\;\raisebox{-1.5mm}{\epsfysize=6mm\epsfbox{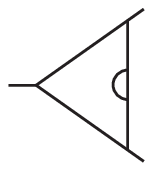}}\;}

\def\w{\;\raisebox{-1.5mm}{\epsfysize=6mm\epsfbox{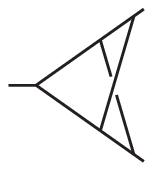}}\;}

\def\p{\;\raisebox{-1.5mm}{\epsfysize=6mm\epsfbox{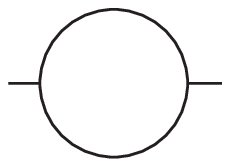}}\;}
\def\pv{\;\raisebox{-1.5mm}{\epsfysize=6mm\epsfbox{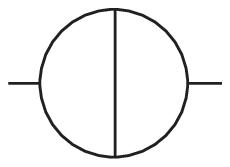}}\;}
\def\pdp{\;\raisebox{-1.5mm}{\epsfysize=6mm\epsfbox{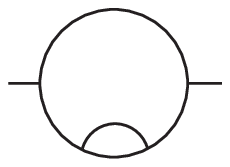}}\;}

\def\vv{\;\raisebox{-1.5mm}{\epsfysize=6mm\epsfbox{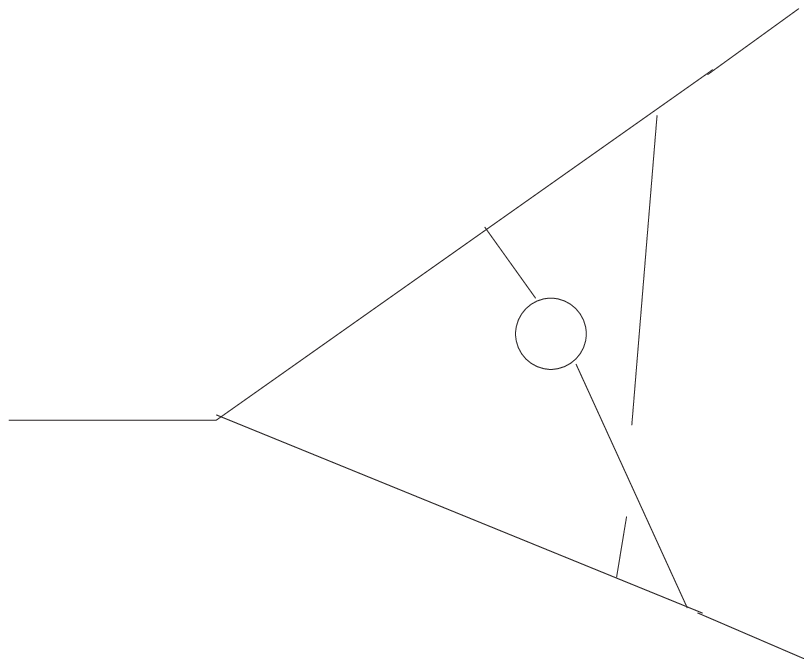}}\;}

\def\vs{\;\raisebox{-1.5mm}{\epsfysize=6mm\epsfbox{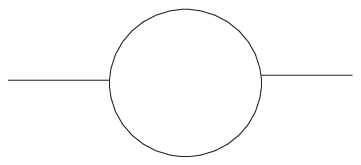}}\;}

\def\vc{\;\raisebox{-1.5mm}{\epsfysize=6mm\epsfbox{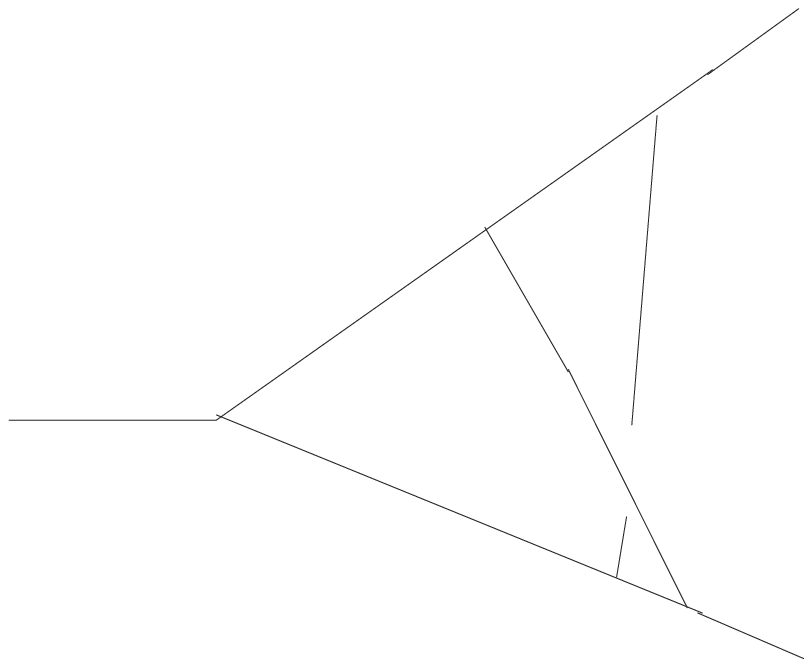}}\;}

\def\vvm{\;\raisebox{-1.5mm}{\epsfysize=6mm\epsfbox{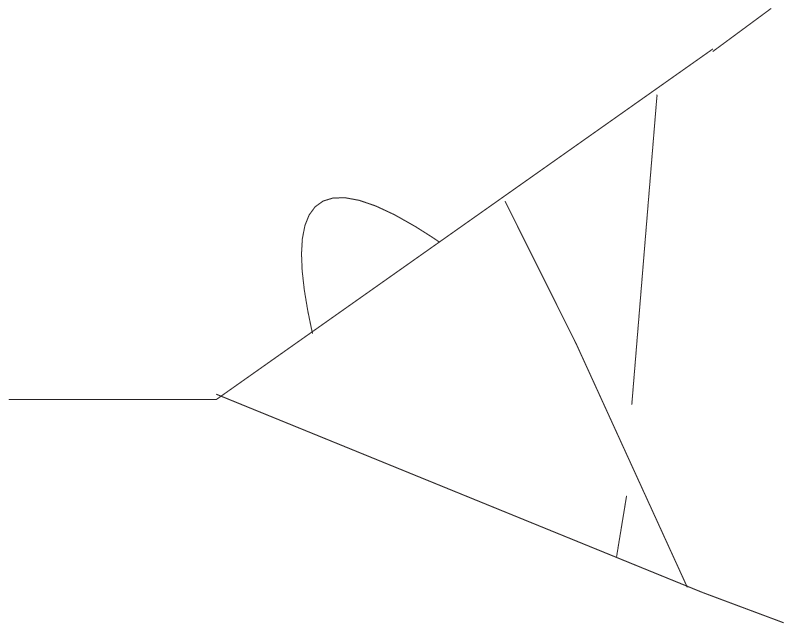}}\;}

\def\se{\;\raisebox{-1.5mm}{\epsfysize=6mm\epsfbox{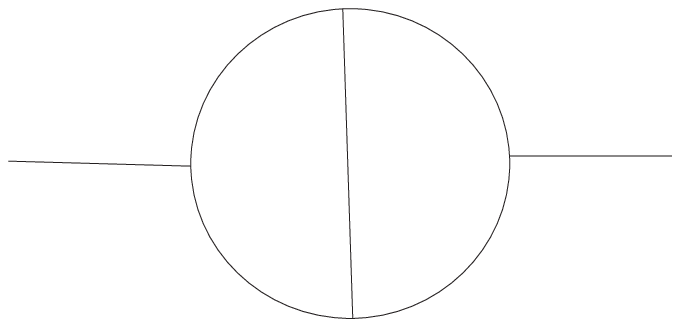}}\;}

\newcommand{\be}{\begin{equation}}
\newcommand{\ee}{\end{equation}}
\newcommand{\bea}{\begin{eqnarray}}
\newcommand{\eea}{\end{eqnarray}}
\newcommand{\beas}{\begin{eqnarray*}}
\newcommand{\eeas}{\end{eqnarray*}}
\newcommand{\bookfig}[5]{
\begin{figure}\centering\fbox{\epsfysize=#5cm \epsfbox{#1}}
\caption[#2]{\small #4}\label{#3}
\end{figure}
}
\begin{document}
\title{Combinatorics of (perturbative) Quantum Field Theory\footnote{hep-th/0010059, MZ-TH/00-42}}
\author{D.~KREIMER\thanks{Heisenberg
Fellow at Mainz Univ., D-55099 Mainz, Germany,
kreimer@thep.physik.uni-mainz.de}\\Lyman Lab., Harvard University}
\date{September 2000}
\maketitle
\begin{abstract}
We review the  structures imposed on perturbative QFT by the fact
that its Feynman diagrams provide  Hopf and Lie algebras. We
emphasize the role which the Hopf algebra plays in renormalization
by providing the forest formulas. We exhibit how the associated
Lie algebra originates from an operadic operation of graph
insertions. Particular emphasis is given to the connection with
the Riemann--Hilbert problem. Finally, we outline how these
structures relate to the numbers which we see in Feynman diagrams.
\end{abstract}
\newpage
\section{Introduction}
Renormalization (see \cite{Collins} for a classical textbook
treatment) has been settled as a self-consistent approach to the
treatment of short-distance singularities in the perturbative
expansion of quantum field theories thanks to the work of
Bogoliubov, Parasuik, Hepp, Zimmermann, and followers.
Nevertheless, its intricate combinatorics went unrecognized for a
long time. In this review we want to describe the results in a
recent series of papers devoted to the Hopf algebra structure of
quantum field theory (QFT)
\cite{DK1,DK2,DKo,CK1,CKl,CKRH,CK2,CK3,BK1,BK2,BK3,DK3,DK4,Book,KD}.
These results were obtained during the last three years, starting
from first papers on the subject \cite{DK1,DK2,DKo} and
flourishing in intense collaborations with Alain Connes
\cite{CK1,CKl,CKRH,CK2,CK3} and David Broadhurst
\cite{BK1,BK2,BK3}.

We will review the results obtained so far in a fairly informative
style, emphasizing the underlying ideas and concepts. Technical
details and mathematical rigor can be found in the above-cited
papers, while it is our present purpose to familiarize the reader
with the key ideas. Furthermore, we intend to spell out lines for
further investigation, as it more and more becomes clear that this
Hopf algebra structure provides a very fine tool for a better
understanding of a correct mathematical formulation of QFT as well
as for applications in particle and statistical physics.

Nevertheless, we will use one concept for the first time in this
paper: we will introduce an operad of Feynman graphs, as it is
underlying many of the operations involved in the Hopf and Lie
algebras built on Feynman graphs.

\section{The Hopf algebra structure: trees and graphs}
Let us start right away with the consideration of how rooted trees
and Feynman graphs are connected in perturbative QFT.
\subsection{Basic considerations}
There are two basic operations on Feynman graphs which govern
their combinatorial structure as well as the process of
renormalization. The question to what extent they also determine
analytic properties of Feynman graphs is one of these future lines
of investigations, with first results in \cite{DK3,DK4}. We will
comment in some detail on this aspect later on.

These two basic operations are the disentanglement of a graph into
subgraphs, and the opposite operation, plugging a subgraph into
another one. Let us consider the disentanglement of a graph first.

We consider the following three-loop vertex-correction $\Gamma$
{\Large $$\Gamma=\vv.$$} We regard it as a contribution to the
perturbative expansion of $\phi^3$ theory in six spacetime
dimensions, where this theory is renormalizable.\footnote{External
lines are amputated, but still drawn, in a convenient abuse of
notation. In the massless case considered here no further notation
is needed for insertions into propagators. In the general case
(massive theories, spin) the external structures defined in
\cite{CK1} are a sufficient tool.} $\Gamma$ contains one
interesting subgraph, the one-loop self-energy graph {\large
$$\gamma=\vs.$$} We are interested in it because it is the only
subgraph which provides a divergence, and the whole UV-singular
structure comes from this subdivergence and from the overall
divergence of $\Gamma$ itself. Let {\large
$$\Gamma_0:=\Gamma/\gamma=\vc$$} be the graph where we shrink
$\gamma$ to a point. From the analytic expressions corresponding
to $\Gamma$, to $ \Gamma_0$ and to $\gamma$ we can form the
analytic expression corresponding to the renormalization of the
graph $\Gamma$. It is given by \be
\Gamma-R(\Gamma)-R(\gamma)\Gamma_0+R\left(R(\gamma)\Gamma_0
\right),\ee where we still abuse, in these introductory remarks,
notation by using the same symbol $\Gamma$ for the graph and the
analytic expression corresponding to it.  We do so as we want to
emphasize for the moment that the crucial step in obtaining this
expression is the use of the graph $\Gamma$ and its disentangled
pieces, $\gamma$ and $\Gamma_0=\Gamma/\gamma$. The analytic
expressions will come as characters on these Hopf algebra
elements, and we will discuss these characters in detail below.
Diagrammatically, the above expression reads {\large $$
\vv-R(\vv)-R(\vs)\vc+R\left(R(\vs)\vc\right).$$}

The unavoidable arbitrariness in the so-obtained expression lies
in the choice of the map $R$ which we suppose to be such that it
does not modify the short-distance singularities (UV divergences)
in the analytic expressions corresponding to the graphs. This then
renders the above combination of four terms finite. If there were
no subgraphs, a simple subtraction $\Gamma-R(\Gamma)$ would
suffice to eliminate the short-distance singularities, but the
necessity to obtain local counterterms forces us to first subtract
subdivergences, which is achieved by Bogoliubov's famous $\bar{R}$
operation \cite{Collins}, which delivers here: \be\Gamma\to
\bar{R}(\Gamma)=\Gamma-R(\gamma)\Gamma_0.\ee This provides two of
the four terms above. Amongst them, these two are free of
subdivergences and hence provide only a local overall divergence.
The projection of these two terms into the range of $R$ provides
the other two terms, which combine to the counterterm \be
Z_\Gamma=-R(\Gamma)+R(R(\gamma)\Gamma_0)\ee of $\Gamma$, and
subtracting them delivers the finite result above by the fact that
the UV divergences are not changed by the renormalization map
$R$.\footnote{Locality is connected to the absence of
subdivergences: if a graph has a sole overall divergence, UV
singularities only appear when all loop momenta tend to infinity
jointly. Regarding the analytic expressions corresponding to a
graph as a Taylor series in external parameters like masses or
momenta, powercounting  establishes that only the coefficients of
the first few polynomials in these parameters are UV singular.
Hence they can be subtracted by a counterterm polynomial in fields
and their derivatives. The argument fails as long as one has not
eliminated all subdivergences: their presence can force each term
in the Taylor series to be divergent.}

The basic operation here is the disentanglement of the graph
$\Gamma$ into pieces $\gamma$ and $\Gamma/\gamma$, and this very
disentanglment gives rise to a Hopf algebra structure, as was
first observed in \cite{DK1}. This Hopf algebra has a role model:
the Hopf algebra of rooted trees. We first want to get an idea
about this universal Hopf algebra after which all the Hopf
algebras of Feynman graphs are modeled.

Consider the two graphs {\large $$\Gamma_1=\vv,\;\Gamma_2=\vvm.$$}
They have one common property: both of them can be regarded as the
graph {\large $$\Gamma_0=\Gamma_1/\gamma=\Gamma_2/\gamma=\vc$$}
into which the subgraph {\large $$\gamma=\vs$$} is inserted, at
two different places though. But as far as their UV-divergent
sectors go they both realize a rooted tree of the form given in
Fig.(\ref{ft}), in the language of \cite{DK1} both graphs
$\Gamma_1,\Gamma_2$ correspond to a parenthesized word of the form
{\large $$((\vs)\vc) .$$} \bookfig{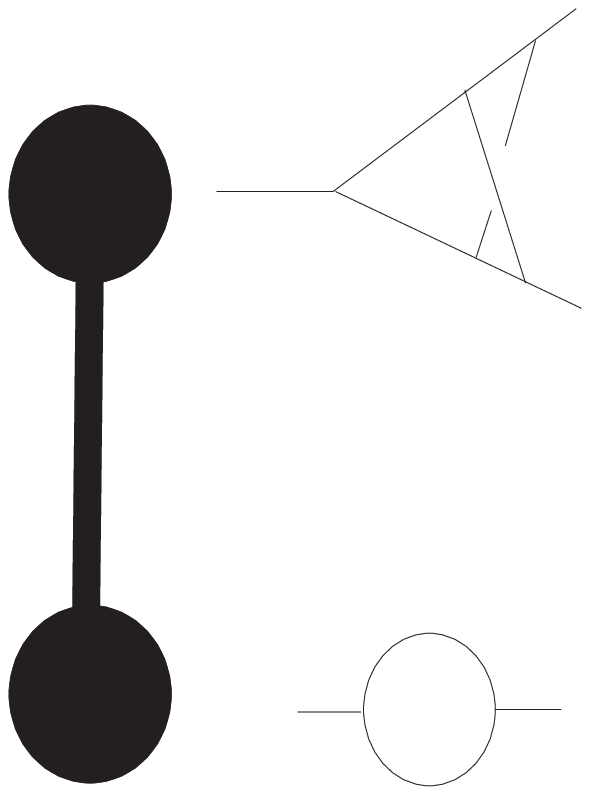}{}{ft}{A decorated
rooted tree with two vertices, each decorated by a graph without
subdivergences (assuming this is an example in $\phi^3$ theory in
six dimensions). The root (by our convention the uppermost vertex)
is decorated by the graph
$\Gamma_0=\Gamma_1/\gamma=\Gamma_2/\gamma$ which we obtain when we
shrink the subdivergence $\gamma$ to a point in either $\Gamma_1$
or $\Gamma_2$ . The vertex decorated by the one-loop self-energy
$\gamma$ corresponds to this subdivergence, and the rooted tree
stores the information that this divergence is nested in the other
graph. The information at which place the subdivergence is to be
inserted is not stored in this notation. The hierarchy which
determines the recursive mechanism of renormalization is
independent of this information. It can easily be restored
allowing marked graphs as decorations, or one could directly
formulate the Hopf algebra on graphs as we do below.}{3} In
\cite{DK1} such graphs were considered to be equivalent, as the
combinatorial process of renormalization produces exactly the same
terms for both of them. We will formulate this equivalence in a
later section using the language of operads.

The combinatorics of renormalization is essentially governed by
this bookkeeping process of the hierarchies of subdivergences, and
this bookkeeping is what is delivered by rooted trees. They are
just the appropriate tool to store the hierarchy of disjoint and
nested subdivergences. Another example given in Fig.(\ref{exa})
might be better suited than any formalism to make this clear.
\bookfig{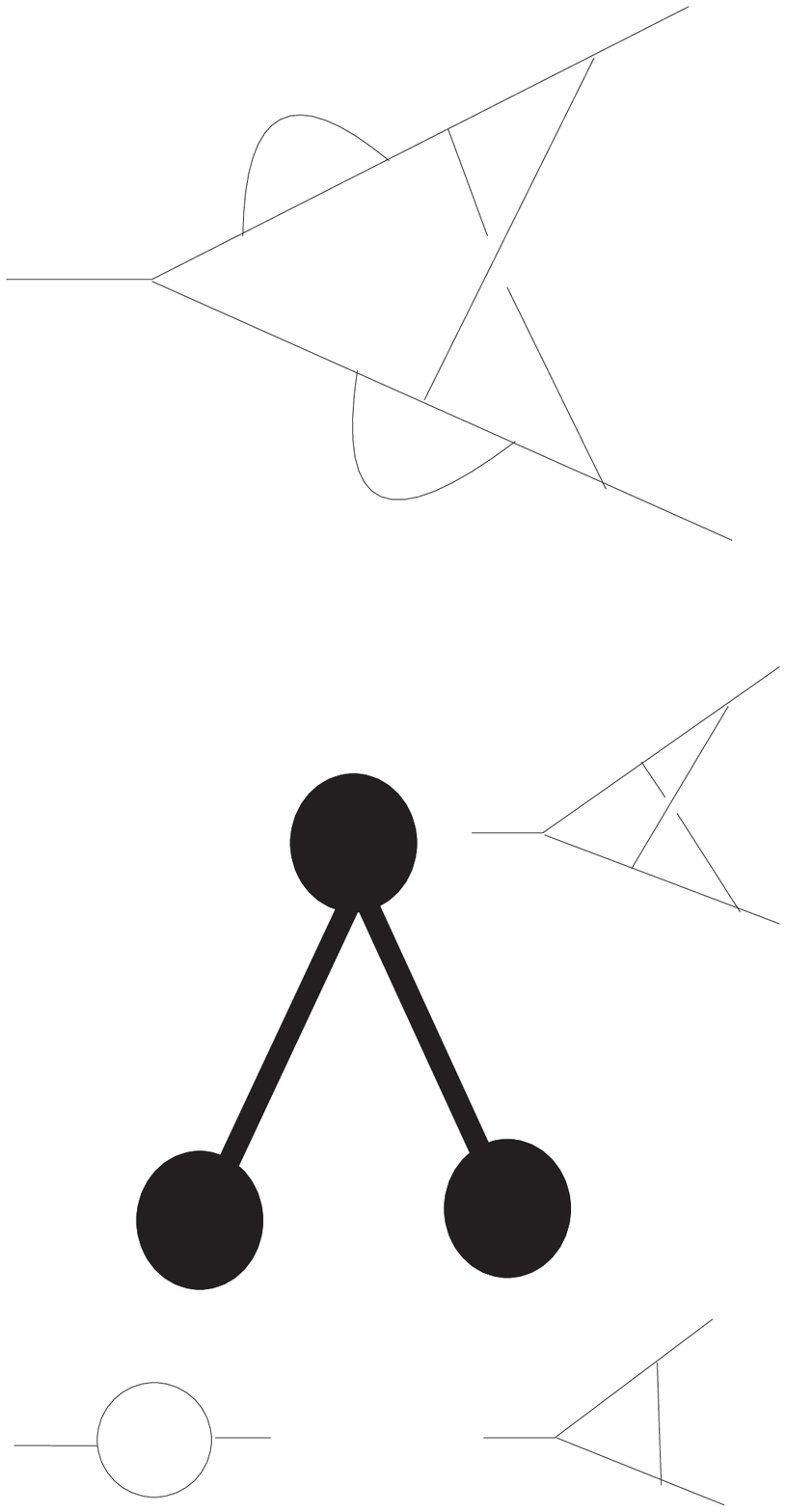}{}{exa}{This graph has a hierarchy of divergences
given by two disjoint subdivergences, the self-energy $\gamma$ and
a one-loop vertex-correction $\tilde{\gamma}$, so that its
divergent structure represents the decorated rooted tree
indicated. As a parenthesized word, the graph corresponds to
$((\gamma)(\tilde{\gamma})\Gamma_0)$. There are, by the way,
$5\times 6=30$ graphs which are all equivalent in the sense that
they represent this rooted tree or parenthesized word, generated
by the 5 internal vertices and 6 internal edges which provide
places for insertion in $\Gamma_0$.}{4}

At this stage, the reader should wonder what to make out of graphs
which have overlapping divergences. This can be best understood
when we turn to the other basic operation on graphs: plugging them
into each other. On the one hand, for the non-overlapping graphs
$\Gamma_1,\Gamma_2$ above there is a unique way to obtain them
from {\large $$\Gamma_0=\Gamma_1/\gamma=\Gamma_2/\gamma=\vc$$} and
the self-energy $\gamma$. We plug $\gamma$ into the
vertex-correction at an appropriate internal line to obtain these
graphs. This operation will be considered in some more detail in a
later section. On the other hand, for a graph which contains
overlapping divergences we have typically no unique manner, but
several ways instead, how to obtain it. For example,
$$\Omega=\se$$ can be obtained in the two ways indicated in
Fig.(\ref{om}). \bookfig{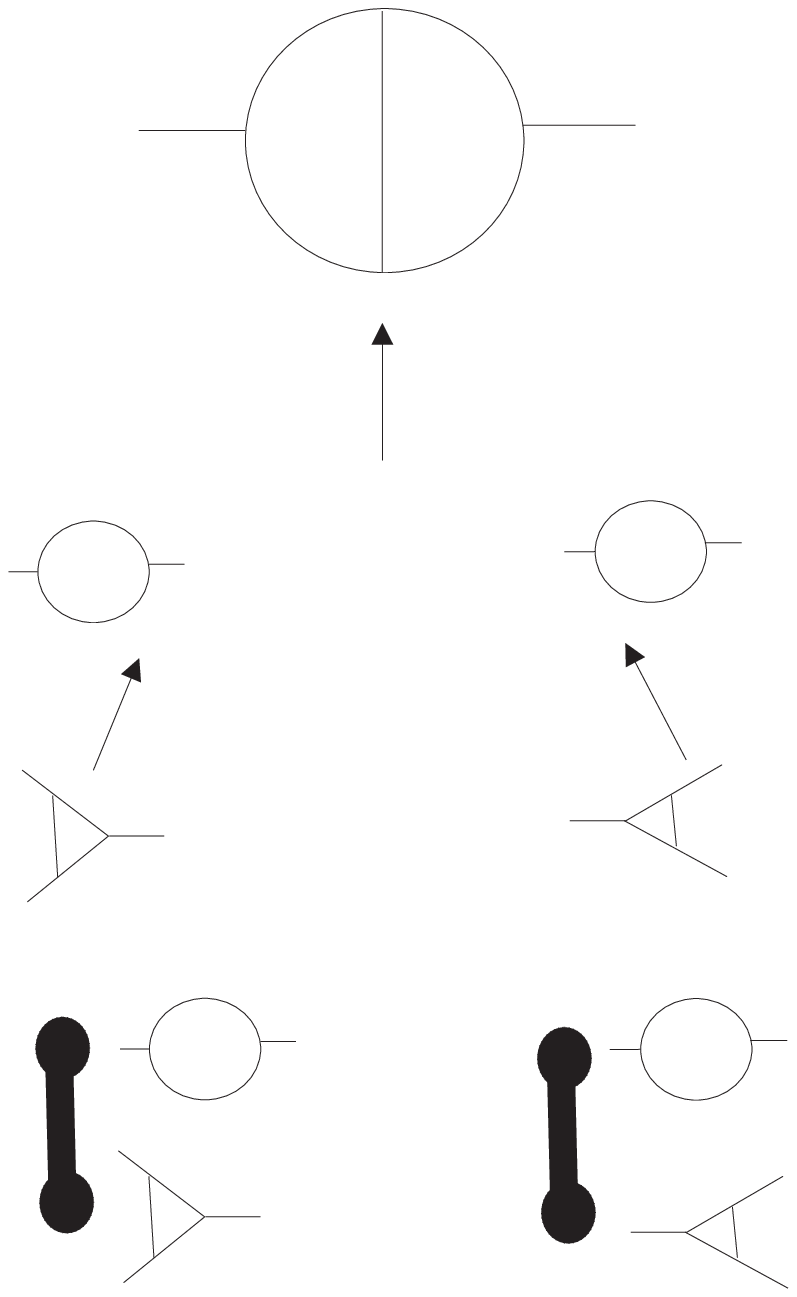}{}{om}{Finding the two ways of
getting the overlapping graph $\Omega$. There are two vertices in
the one-loop self-energy into which the one-loop vertex correction
can be inserted. Both result in the same graph. The short-distance
singularities in $\Omega$ arise from two sectors, described by two
decorated rooted trees.}{5}

Each of these ways corresponds to a rooted tree \cite{DKo}, and
the sum over all these rooted trees bookkeeps the subdivergent
structures of a graph with overlapping divergences correctly. The
resolution of overlapping divergences into rooted trees
corresponds to the determination of Hepp sectors, and amounts to a
resolution of overlapping subsets into nested and disjoint subsets
generally \cite{DKo}.\footnote{The remarks above are specific to
theories which have trivalent couplings. In general, the
determination of divergent sectors still leads to rooted trees
\cite{DKo}. A concrete example how the Hopf algebra structure
appears in $\phi^4$-theory can be found in \cite{GL}. Also,
resolving the overlapping divergences in terms of decorated rooted
trees determines the appropriate set of primitive elements of the
Hopf algebra, which can for example be systematically achieved by
making use of Dyson--Schwinger equations \cite{DK1,Book}, see also
\cite{KW}.}

One remark is in order: the very fact that overlapping divergences
can be reduced to divergences which have a tree-hierarchy has a
deeper reason: the short-distance singularities of QFT result from
confronting  products of distributions which are well-defined on
the configuration space of vertices located at distinct space-time
points, but which become ill-defined along diagonals \cite{EG,S}.
But then, the various possible ways how an ensemble of distinct
points can collapse to (sub-)diagonals is known to be stratified
by rooted trees \cite{FulMacPh}, and this is what essentially
ensures that the Hopf algebra structure of these trees can
reproduce the forest formulas of perturbative QFT. Let us then
have a closer look at the connection between graphs and rooted
trees.
\subsection{Sector decomposition and rooted trees}
Consider the Feynman graph $\Omega$ once more, as given in
Fig.(\ref{f1}).  It corresponds to a contribution to the
perturbative expansion in the coupling constant $g$ of the theory
to order $g^4$. It has short-distance (UV) singularities which are
apparent in the following sectors $$ I_1:=\{1,2,3 \},I_2:=\{2,3,4
\}, I=\{1,2,3,4\},$$ which give the label of the vertices
participating in the divergent (sub-)\-graphs.
 Note that the
sectors overlap: $I_1\cap I_2\not=\emptyset$. The singularities
are stratified so that they can be represented as rooted trees, as
described in Fig.(\ref{f1}). In this stratification of sectors
each node at the rooted tree corresponds to a Feynman graph which
connects the vertices attached to the node by propagators in a
manner such that it has no subdivergences. We call such graphs
primitive graphs. Each primitive graph is only overall divergent.
\bookfig{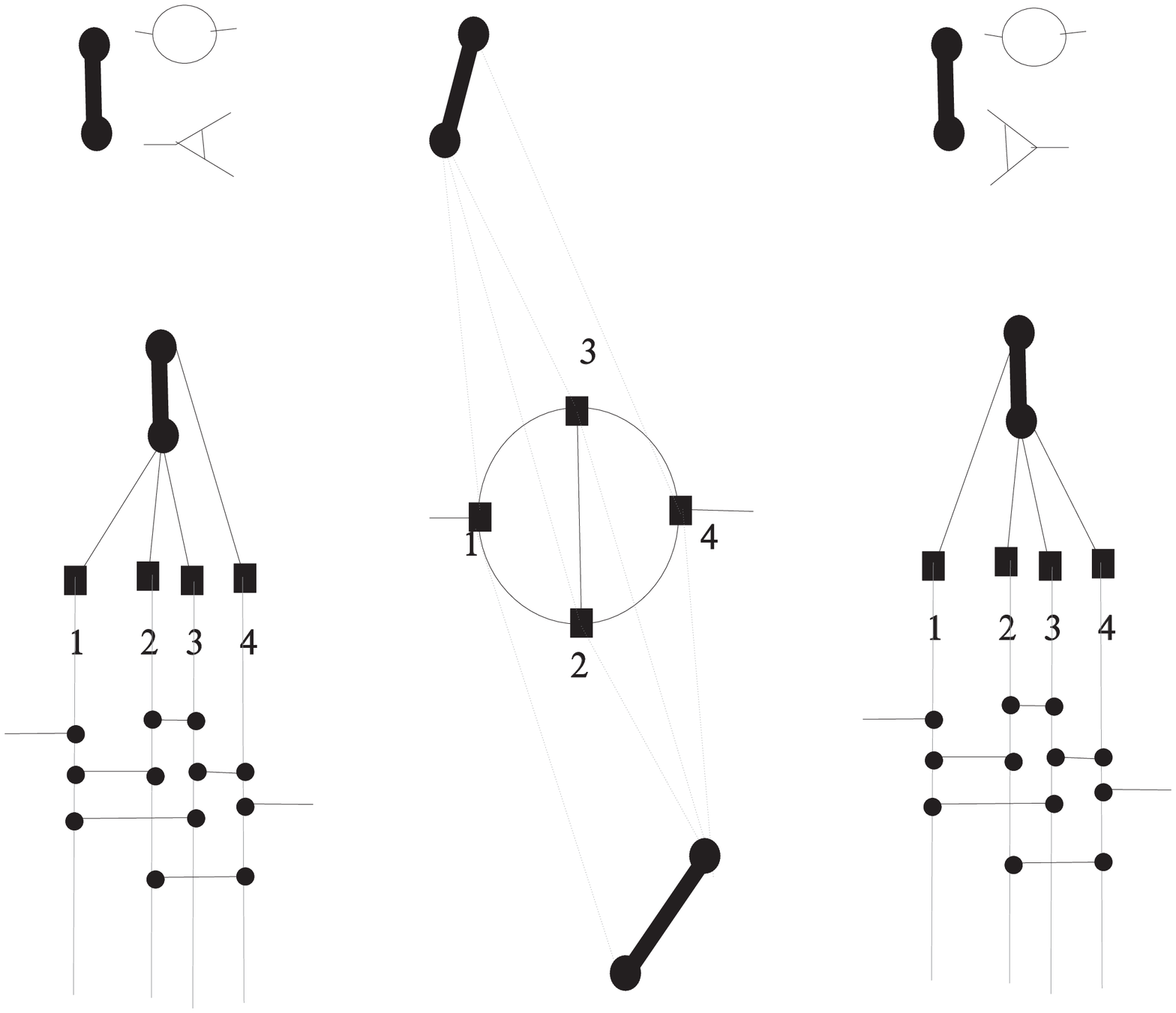}{}{f1}{A different way of looking at the graph
$\Omega$. We label its vertices by $1,\ldots,4$. Then, the set of
vertices $\{1,2,3\}$ belongs to a vertex subgraph, as does the set
$\{2,3,4\}$. The fact that both are proper subsets of the set of
all vertices $\{1,2,3,4\}$ is again reflected in a tree-like
hierarchy. A short distance singularity appears when these
labelled vertices of a divergent graph collapse to a point. This
point (a diagonal) constitutes one vertex of a rooted tree, with
the root corresponding to the collapse of all vertices to the
overall diagonal jointly. Again, the graph gives rise to two
rooted trees, which corresponds to the two divergent sectors along
two different diagonals. When we blow up the vertices
$\{1,2,3,4\}$ of the graph to vertical lines, we can represent the
edges of the Feynman graph as horizontal chords, and we regain the
graph by shrinking the vertical lines to a point. The Hopf algebra
structure operates on the bold black rooted trees, as they store
the information which diagonals contain short-distance
singularities in the graph under consideration.}{9}

Now, where do singularities reside? Typically, if we write down
analytic expressions in terms of momentum integrals,
UV-divergences appear when the loop momenta involved in a
primitive graph tend to infinity jointly, and this can be detected
by powercounting over edges and vertices in the graph. On the
other hand, we can consider Feynman rules in coordinate space.
Then, the UV-singular integrations over momenta become short
distance singularities. Again, they creep in from the very fact
that closed loops, cycles in the graph, force the integration over
the positions of vertices to produce ill-defined products of
distributions with coinciding support. Powercounting amounts to a
check of the scaling degree of the relevant distributions and
ultimately determines the appearance of a short distance
singularity at the diagonal under consideration.

The short distance singularities of Feynman graphs then come
solely from  regions where all vertices are located at coinciding
points. One has no problem to define the Feynman integrand in the
configuration space of vertices at distinct locations, while a
proper extension to diagonals is what is required.

In the above, the two divergent subgraphs are ill-defined along
the diagonals $x_1=x_2=x_3$  and $x_2=x_3=x_4$  while the overall
divergence corresponds to the main diagonal $x_1=x_2=x_3=x_4$.

Due to the Hopf algebra structure of Feynman graphs we can define
the renormalization of all such sectors without making recourse to
any specific analytic properties of the expressions (Feynman
integrals) representing those sectors. The only assumption we make
is that in a sufficiently small neighborhood of such an ultralocal
region (the neighborhood of a diagonal) we can define the scaling
degree, --the powercounting--, in a sensible manner.

Apart from this assumption our approach is purely combinatorial
and in particular independent of the geometry of the underlying
spacetime manifold.

Fig.(\ref{f1}) also gives a first idea why the Hopf algebra of
undecorated rooted trees is the universal object underlying the
Hopf algebras of Feynman graphs. The essential combinatorics
needed to obtain local counterterms will solely use cuts on these
rooted trees which are drawn in bold black lines in the figure,
with no further operation on decorations. Different theories just
differ by having different types of chords and vertices, while to
each chord and vertex in the figure we assign the appropriate
scaling degree, the weight with which they contribute to the
powercounting.

One further remark is in order: the existence of a purely
combinatorial solution coincides with the result of Brunetti and
Fredenhagen \cite{BF}, who showed that the renormalization
mechansism is indeed unchanged in the context of curved manifolds
in a detailed local analysis using the Epstein--Glaser mechanism.
To my mind, quite generally, the Hopf algebra can be used to make
sense out of extensions of products of distributions to diagonals
of configuration spaces even before we decide by which class of
(generalized) functions we want to realize these extensions. While
consistency of the Hopf algebra approach to renormalization with
the Epstein--Glaser formalism was settled once the Hopf algebra
was directly formulated on graphs \cite{DKo,CK2}, it was also
addressed at a notational level making use of configuration space
Feynman graphs in \cite{GL}. Still, one should regard the
splitting of distributions itself as the first instance where a
representation of the Hopf algebra is realized, so that properties
like Lorentz covariance appear as properties of the representation
alone, maintaining a proper separation of the combinatorics of the
Bogoliubov recursion from the analytic properties of the functions
defined over the configuration space, enabling also a direct
formulation on the level of time-ordered products instead of
Feynman graphs.

Once more, that the Hopf algebra structure coming in is the one of
rooted trees should be no surprise: limits to diagonals in
configuration spaces are stratified by rooted trees
\cite{FulMacPh}, and it is the Hopf algebra structure of these
rooted trees which describes the combinatorics of renormalization,
as we will see. The Hopf algebra of rooted trees will be the role
model for all the Hopf algebras of Feynman graphs for a
specifically chosen QFT, a classifying space in technical terms
(see Theorem 2, section 3 in \cite{CK1}), while each such chosen
QFT probes the short distance singularities according to its
Feynman graphs. The resulting iterative procedure gives rise to
the Hopf algebra of rooted trees which was first described, in the
equivalent language of parenthesized words, in \cite{DK1} and then
in its final notation in \cite{CK1}. It is now time to describe
this Hopf algebra of rooted trees in some detail.
\subsection{The Hopf algebra of undecorated rooted trees}
We follow section II of \cite{CK1} closely. A {\em rooted tree}
$t$ is a connected and simply-connected set of oriented edges and
vertices such that there is precisely one distinguished vertex
which has no incoming edge. This vertex is called the root of $t$.
Further, every edge connects two vertices and the {\em fertility}
$f(v)$ of a vertex $v$ is the number of edges outgoing from $v$.
The trees being simply-connected, each vertex apart from the root
has a single incoming edge (we could attach, if we like, an extra
edge to the root as well, for a more common treatment). Each
vertex in such a rooted tree corresponds to a divergent sector in
a Feynman diagram. The rooted trees store the hierarchy of such
sectors. We will always draw the root as the uppermost vertex in
figures, and agree that all edges are oriented away from the root.

%%XXX \special{src: 101 OVERL.TEX} %Inserted by TeXtelmExtel

As in \cite{CK1}, we consider the (commutative) algebra of
polynomials over $\Qb$ in rooted trees, where the multiplication
$m(t,t^\prime)$ of two rooted trees is their disjoint union, so we
can draw them next to each other in arbitrary order, and the unit
with respect to this multiplication is the empty set.\footnote{We
restrict ourselves to one-particle irreducible diagrams for the
moment. Then, the disjoint union of trees corresponds to the
disjoint union of graphs. One could also set up the Hopf algebra
structure such that one-particle reducible graphs correspond to
products of rooted trees \cite{DK1}.} Note that for any rooted
tree $t$ with root $r$ which has fertility $f(r)=n\geq 0$, we have
trees $t_1$, $\ldots$, $t_n$ which are the trees attached to $r$.

Let $B_-$ be the operator which removes the root $r$ from a tree
$t$, as in Fig.(\ref{guill-}):
\begin{equation}
B_-: t\to B_-(t)=t_1 t_2\ldots t_n.
\end{equation}
\bookfig{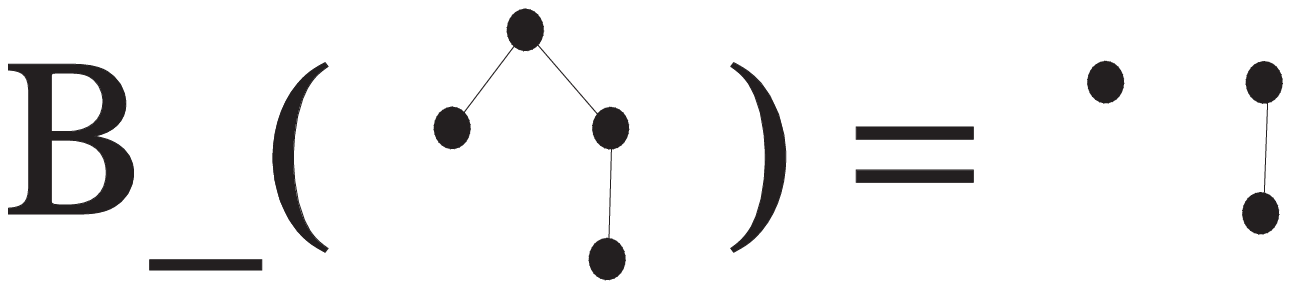}{$B_-$}{guill-}{The action of $B_-$ on an
undecorated rooted tree.}{1} We extend the action of $B_-$ to a
product of rooted trees by a Leibniz rule,
$B_-(XY)=B_-(X)Y+XB_-(Y)$. We also set $B_-(t_1)=1$, $B_+(1)=t_1$,
where $t_1$ is the rooted tree corresponding to the root alone.

%%XXX \special{src: 130 OVERL.TEX} %Inserted by TeXtelmExtel

Let $B_+$ be the operation which maps a monomial of $n$ rooted
trees to a new rooted tree $t$ which has a root $r$ with fertility
$n$ which connects to the $n$ roots of $t_1,\ldots,t_n$:
\begin{equation}
B_+: t_1\ldots t_n\to B_+(t_1\ldots t_n)=t.
\end{equation}
This is clearly the inverse to the action of $B_-$ on single
rooted trees. One has
\begin{equation}
B_+(B_-(t))=B_-(B_+(t))=t
\end{equation}
for any rooted tree $t$. Fig.(\ref{guill+}) gives an example.
\bookfig{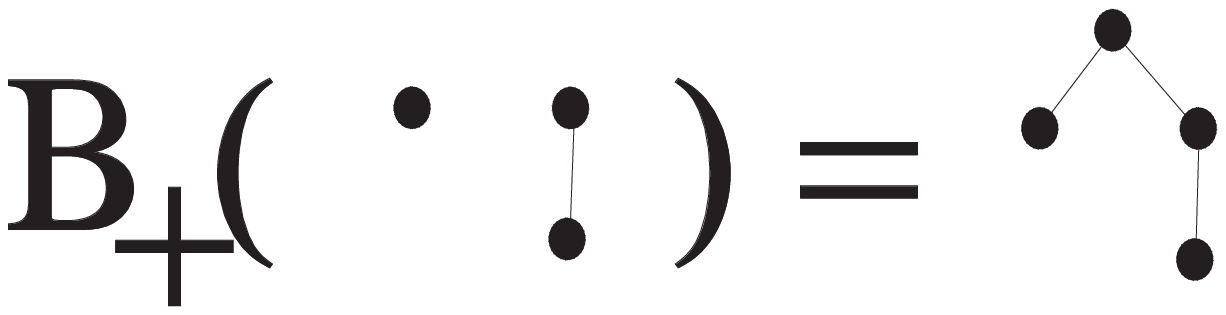}{$B_+$}{guill+}{The action of $B_+$ on a monomial
of trees.}{1}

All the operations described here have a straightforward
generalization to decorated rooted trees, in which case the
operator $B_+$ carries a further label to indicate the decoration
of the root \cite{CK1}. We will not use decorated rooted trees
later, as we will directly formulate the Hopf algebras of specific
QFTs on Feynman graphs. The Hopf algebra of undecorated rooted
trees is the universal object \cite{CK1} for all those Hopf
algebras, and hence we describe it here in some detail.

%%XXX \special{src: 141 OVERL.TEX} %Inserted by TeXtelmExtel

%%XXX \special{src: 146 OVERL.TEX} %Inserted by TeXtelmExtel

 Note
that while $[B_+,B_-](t)=0$ for any single rooted tree $t$, this
commutator is non-vanishing on products of trees. Obviously, one
always has ${\rm id}=B_-B_+$, while $B_+B_-$ acts trivially only
on single rooted trees, not on their product.\footnote{This has
far reaching consequences and is closely connected to the fact
that logarithmic derivatives (with respect to the log of some
scale say) of $Z$-factors are finite quantities. Indeed,
$Z$-factors can be regarded as formal series over  Feynman
diagrams graded by the loop number starting with 1, and their
logarithm defines a series in graphs which typically demands that
commutators like $[B_+,B_-](t_1 t_1)$ are a primitive element in
the Hopf algebra, and hence provide only a first order pole
\cite{KD,BK3}. This is a first instance of a t'Hooft relation to
which we turn later when we review the results of \cite{CK3}.}

We will introduce a Hopf algebra on our rooted trees by using the
possibility to cut such trees in pieces. For the reader not
familiar with Hopf algebras, let us mention a few very elementary
facts first. An algebra $A$ is essentially specified by a binary
operation $m:A\times A\to A$ (the product) fulfilling the
associativity $m(m(a,b),c)=m(a,m(b,c))$ so that to each two
elements of the algebra we can associate a new element in the
algebra, and by providing some number field ${\Kb}$ imbedded in
the algebra via $E:\Kb\to A$, $k\to k1$. In a coalgebra we do the
opposite, we disentangle each algebra element: each element $a$ is
decomposed by the coproduct $\Delta:A\to A\times A$ in a
coaasociative manner, $(\Delta\times {\rm id})\Delta(a)=({\rm
id}\times \Delta)\Delta(a)$. Further, the unit $1$ of the algebra,
$m(1,a)=m(a,1)=a$, is dualized to the counit $\bar{e}$ in the
coalgebra, $(\bar{e}\times {\rm id})\Delta(a)=({\rm id}\times
\bar{e})\Delta(a)=a$. If the two operations $m,\Delta$ are
compatible (the coproduct of a product is the product of the
coproducts), we have a bialgebra, and if there is a coinverse, the
celebrated antipode $S:H\to H$, as well, we have a Hopf algebra.
While in the algebra the unit, the inverse and the product are
related by $m(a,a^{-1})=m(a^{-1},a)=1$, the counit, the coproduct
and the coinverse are related by $m(S\times {\rm id})\Delta=E\circ
\bar{e}$. A thorough introduction can be found for example in
\cite{Kassel}.

To define a coproduct for rooted trees we are hence looking for a
map which disentangles rooted trees. We start with the most
elementary possibility. An {\em elementary cut} is a cut of a
rooted tree at a single chosen edge, as indicated in
Fig.(\ref{ecut}). By such a cutting procedure, we will obtain the
possibility to define a coproduct, as we can use the resulting
pieces on either side of the coproduct. It is this cutting
operation which corresponds to the disentanglements of graphs
discussed before. \bookfig{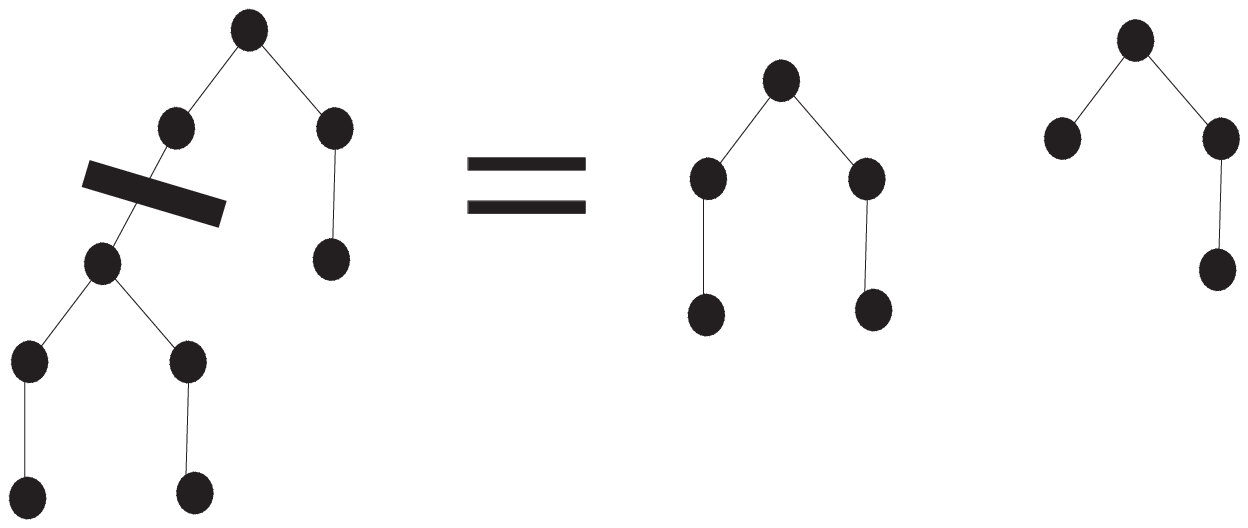}{Elementary cut.}{ecut}{An
elementary cut $c$ splits a rooted tree $t$ into two components.
We remove the chosen edge and get two components. Both are rooted
trees in an obvious manner: one contains the vertex which was the
old root and the root of the other is provided by the vertex which
was at the endpoint (edges are oriented away from the root) of the
removed edge.}{2}

%%XXX \special{src: 179 OVERL.TEX} %Inserted by TeXtelmExtel

Still before introducing the coproduct we introduce the notion of
an {\em admissible cut}, also called a {\em simple cut}
\cite{CK1}. It is any assignment of elementary cuts to a rooted
tree $t$ such that any path from any vertex of the tree to the
root has at most one elementary cut, as in  Fig.(\ref{cut}). An
admissible cut $C$ maps a tree to a monomial in trees. If the cut
$C$ contains $n$ elementary cuts, it induces a map
\begin{equation}
C: t\to C(t)=\prod_{i=1}^{n+1} t_{j_i}.
\end{equation}
Note that precisely one of these trees $t_{j_i}$ will contain the
root of $t$. Let us denote this distinguished tree by $R^C(t)$.
The monomial which is delivered by the $n-1$ other factors is
denoted by $P^C(t)$. In graphs, $P^C(t)$ corresponds to a set of
disjoint subgraphs $\cup_i\gamma_i$ which we shrink to a point and
take out of the initial graph $\Gamma$ corresponding to $t$, while
$R^C(t)$ corresponds to the remaining graph
$\Gamma/(\cup_i\gamma_i)$. Admissibility means that there are no
further disentanglements in the set $\cup_i\gamma_i$. Hence, a sum
over all such sets provides a sum over all unions of subgraphs, as
we will discuss below. Arbitrary non-admissible cuts correspond to
the notion of forests in the sense of Zimmermann \cite{DK1,CK1}.
\bookfig{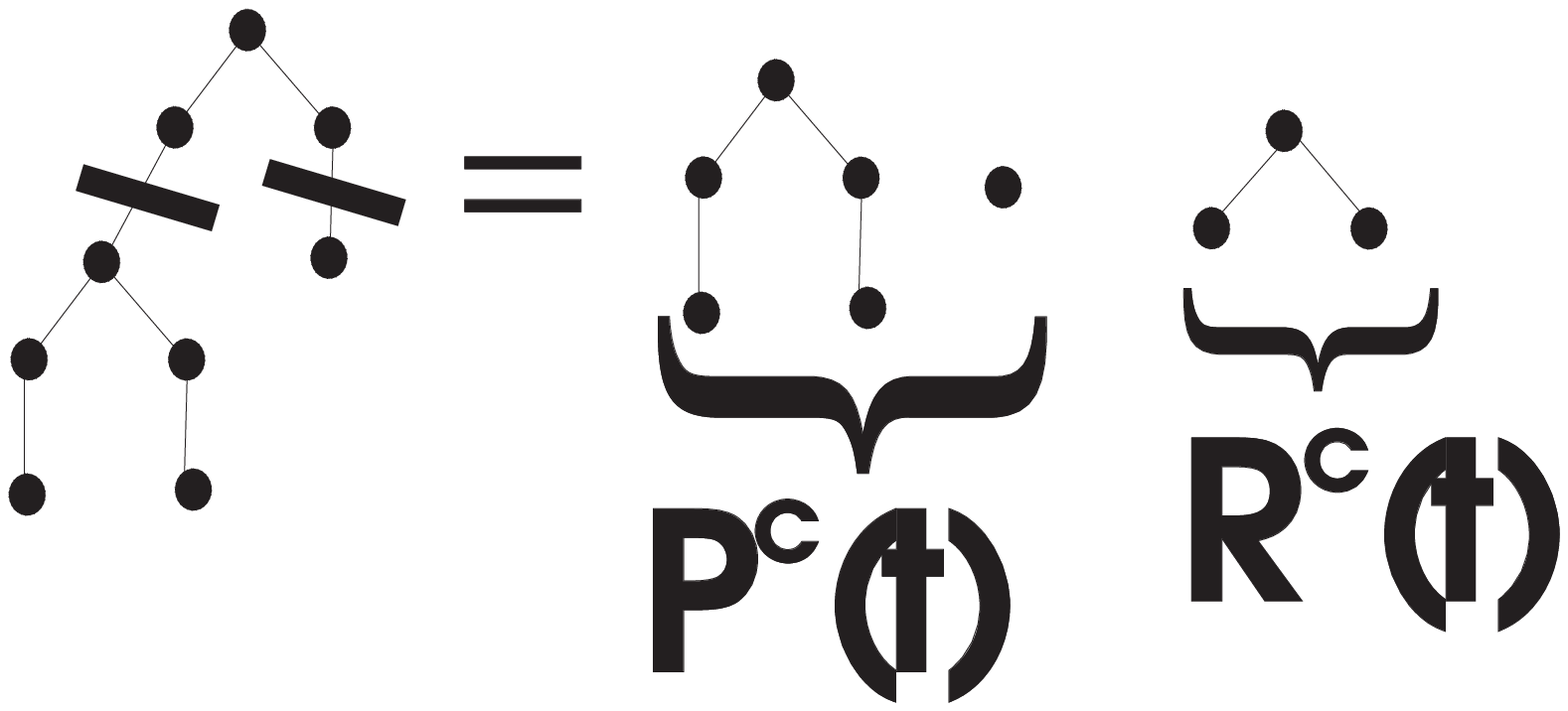}{An admissible cut.}{cut}{An admissible cut $C$
acting on a tree $t$. It produces a monomial of trees. One of the
factors, $R^C(t)$, contains the root of $t$.}{2.7}

%%XXX \special{src: 194 OVERL.TEX} %Inserted by TeXtelmExtel

%%XXX \special{src: 211 OVERL.TEX} %Inserted by TeXtelmExtel

%The definitions of $C,P,R$ can be extended to monomials of trees
%in the obvious manner, by choosing a cut $C^i$ for every tree
%$t_{j_i}$ in the monomial:
%\begin{eqnarray*}
%C(t_{j_1}\ldots t_{j_n}) & := & C^1(t_{j_1})\ldots C^n(t_{j_n}),\\
%P^C(t_{j_1}\ldots t_{j_n}) & := & P^{C^1}(t_{j_1})\ldots
%P^{C^n}(t_{j_n}),\\ R^C(t_{j_1}\ldots t_{j_n}) & := &
%R^{C^1}(t_{j_1})\ldots R^{C^n}(t_{j_n}).
%\end{eqnarray*}

%%XXX \special{src: 223 OVERL.TEX} %Inserted by TeXtelmExtel

%%XXX \special{src: 226 OVERL.TEX} %Inserted by TeXtelmExtel

%%XXX \special{src: 229 OVERL.TEX} %Inserted by TeXtelmExtel

Let us now establish the Hopf algebra structure. Following
\cite{DK1,CK1} we define the counit and the coproduct. The {\em
counit} $\bar{e}$: $H \to \Qb$ is simple: $$ \bar{e}(X)=0 $$ for
any $X\not= 1$, $$ \bar{e}(1)=1. $$

%%XXX \special{src: 244 OVERL.TEX} %Inserted by TeXtelmExtel

The {\em coproduct} $\Delta$ is defined by the equations
\begin{eqnarray}
\Delta(1) & = & 1\otimes 1,\nonumber\\ \Delta(t_1\ldots t_n) & = &
\Delta(t_1)\ldots \Delta(t_n),\nonumber\\ \Delta(t) & = & t
\otimes 1 +(id\otimes B_+)[\Delta(B_-(t))],\label{cop2}
\end{eqnarray}
which defines the coproduct on trees with $n$ vertices iteratively
through the coproduct on trees with a lesser number of vertices.
\bookfig{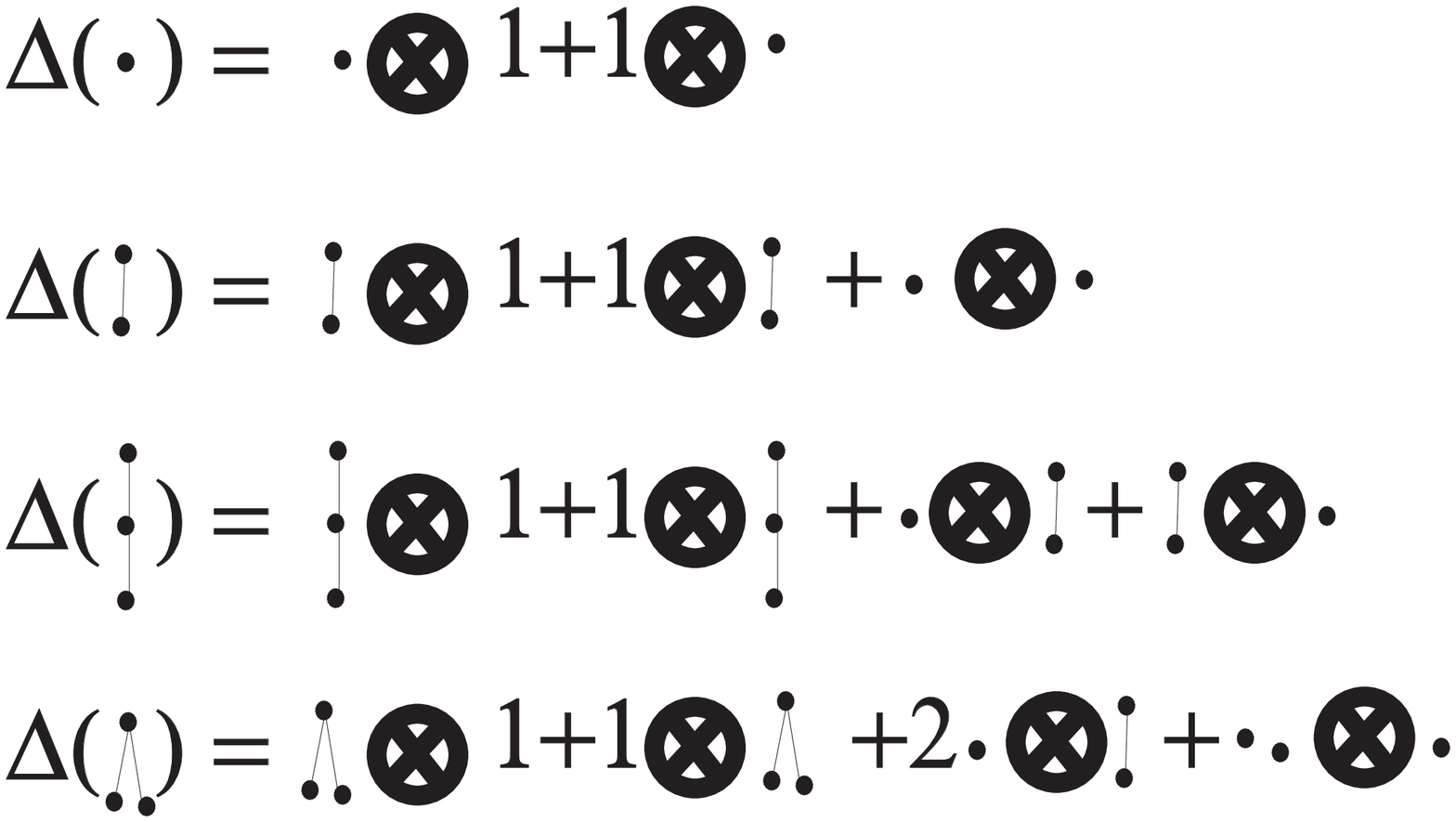}{The coproduct.}{cop}{The coproduct. We work it
out for the trees $t_1,t_2,t_{3_1},t_{3_2}$, from top to
bottom.}{4}

%%XXX \special{src: 269 OVERL.TEX} %Inserted by TeXtelmExtel

The coproduct can be written in a non-recursive manner as
\cite{DK1,CK1}
\begin{equation}
\Delta(t)=1\otimes t+ t\otimes 1+ \sum_{\mbox{\tiny adm.~cuts $C$
of $t$}}P^{C}(t)\otimes R^C(t).\label{cop1}
\end{equation}

%%XXX \special{src: 277 OVERL.TEX} %Inserted by TeXtelmExtel

%%XXX \special{src: 280 OVERL.TEX} %Inserted by TeXtelmExtel

%%XXX \special{src: 283 OVERL.TEX} %Inserted by TeXtelmExtel

%%XXX \special{src: 286 OVERL.TEX} %Inserted by TeXtelmExtel

%%XXX \special{src: 289 OVERL.TEX} %Inserted by TeXtelmExtel

Up to now we have established a bialgebra structure. It is
actually a Hopf algebra. Following \cite{DK1,CK1} we find the
antipode $S$ as
\begin{eqnarray}
S(1) & = & 1,\nonumber\\ S(t_1\ldots t_k) & = & S(t_1)\ldots
S(t_k),\nonumber\\S(t) & = & -t-\sum_{\mbox{\tiny adm.~cuts $C$ of
$t$}}S[P^C(t)]R^C(t).
\end{eqnarray}
Fig.(\ref{ant}) gives examples for the antipode.
\bookfig{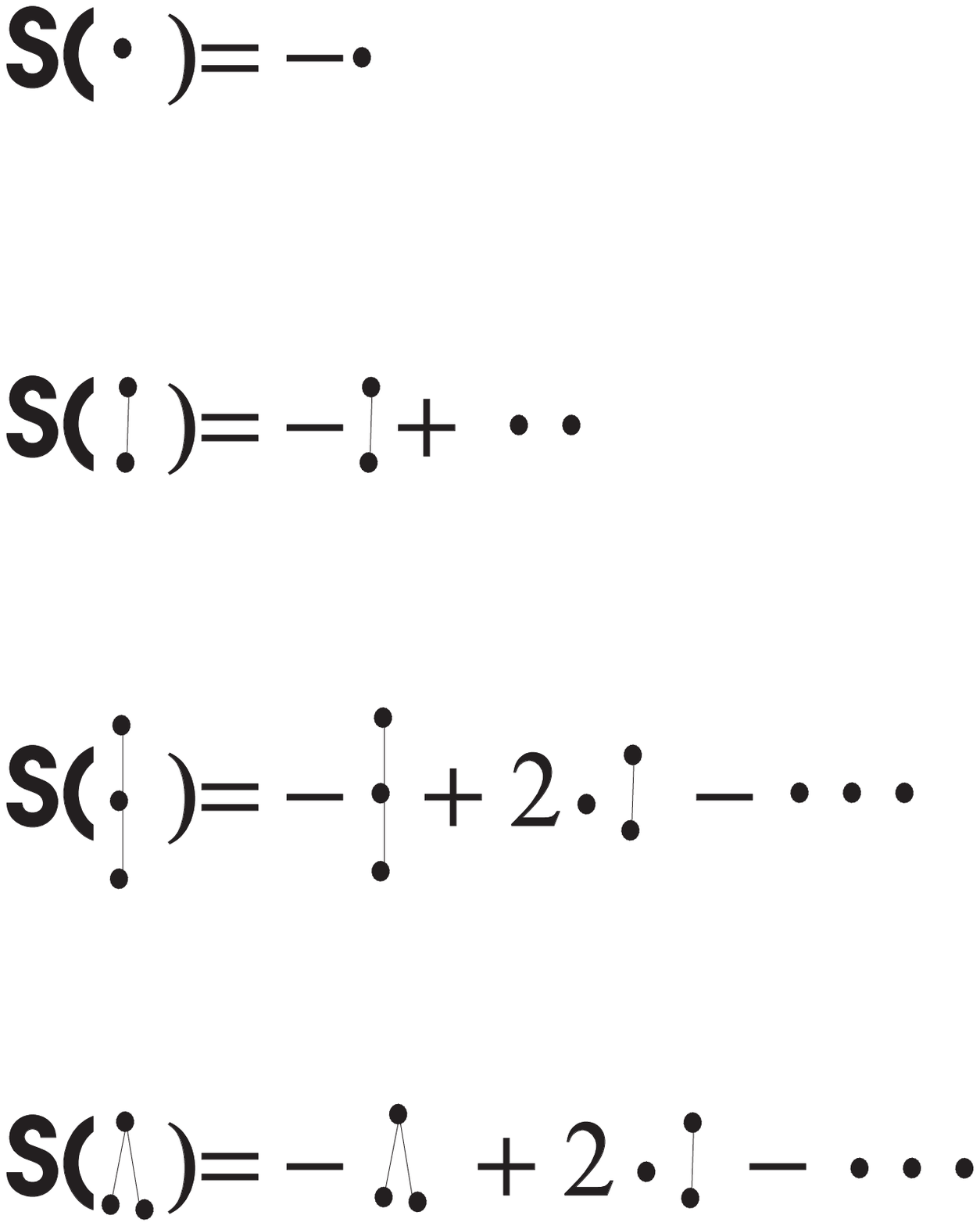}{The coproduct.}{ant}{The antipode. Again we work
it out for the trees $t_1,t_2,t_{3_1},t_{3_2}$.}{5}

%%XXX \special{src: 298 OVERL.TEX} %Inserted by TeXtelmExtel

%%XXX \special{src: 301 OVERL.TEX} %Inserted by TeXtelmExtel

%%XXX \special{src: 304 OVERL.TEX} %Inserted by TeXtelmExtel

Let us give yet another formula to write the antipode, which one
easily derives using induction on the number of vertices
\cite{DK1,CK1}:
\begin{eqnarray}
S(t) & = & -\sum_{\mbox{\tiny all  cuts $C$ of
$t$}}(-1)^{n_C}P^C(t)R^C(t),
\end{eqnarray}
where $n_C$ is the number of elementary cuts in $C$. This time, we
have a non-recursive expression, summing over all cuts $C$,
relaxing the restriction to admissible cuts.

%%XXX \special{src: 319 OVERL.TEX} %Inserted by TeXtelmExtel

%%XXX \special{src: 322 OVERL.TEX} %Inserted by TeXtelmExtel

By now we have established a Hopf algebra $H$ on rooted trees,
using the set of rooted trees, the commutative multiplication $m$
for elements of this set, the unit $1$ and counit $\bar{e}$, the
coproduct $\Delta$ and antipode $S$.  Still following
\cite{DK1,CK1} we allow to label the vertices of rooted trees by
Feynman graphs without subdivergences, in the sense described
before. Quite general, if $Y$ is a set of primitive elements
providing labels, we get a similar Hopf algebra $ H(Y)$. The
determination of all primitive graphs which can appear as labels
corresponds to a skeleton expansion and is discussed in detail in
\cite{DKo}. Instead of using the language of a decorated Hopf
algebra we use directly the corresponding Hopf algebra of graphs
below.

Let us also mention again that
\begin{equation}
m[(S\otimes {\rm id})\Delta(t)]=E\circ
\bar{e}(t)\;\;(=0\;\mbox{for any non-trivial
$t\not=1$}).\label{basic}
\end{equation}
As the divergent sectors in Feynman graphs are stratified by
rooted trees, we can use the Hopf algebra structure to describe
the disentanglement of graphs into pieces, and it turns out that
this delivers the forest formulas of renormalization theory.

%%XXX \special{src: 347 OVERL.TEX} %Inserted by TeXtelmExtel

%%XXX \special{src: 350 OVERL.TEX} %Inserted by TeXtelmExtel

Let us now come back to the graph $\Omega$ and its representation
in Fig.(\ref{f1}). We want to look at the relevant Hopf algebra
operations in some detail, which we describe in Fig.(\ref{f3}).
The operations described in this figure go through for any QFT
whose ultraviolet divergences are local, stratified by rooted
trees that is. A renormalizable field theory will only demand a
finite number of counterterms in the action, while an effective
theory is finite in the number of needed counterterms only for a
finite loop order, but the number will actually increase with the
loop order. A superrenormalizable theory gives only a truncated
representation of rooted trees: higher orders in the perturbative
expansion do not deliver new short-distance singularities, and
hence the existent divergences are stratified by rooted trees with
a restricted number of vertices. \bookfig{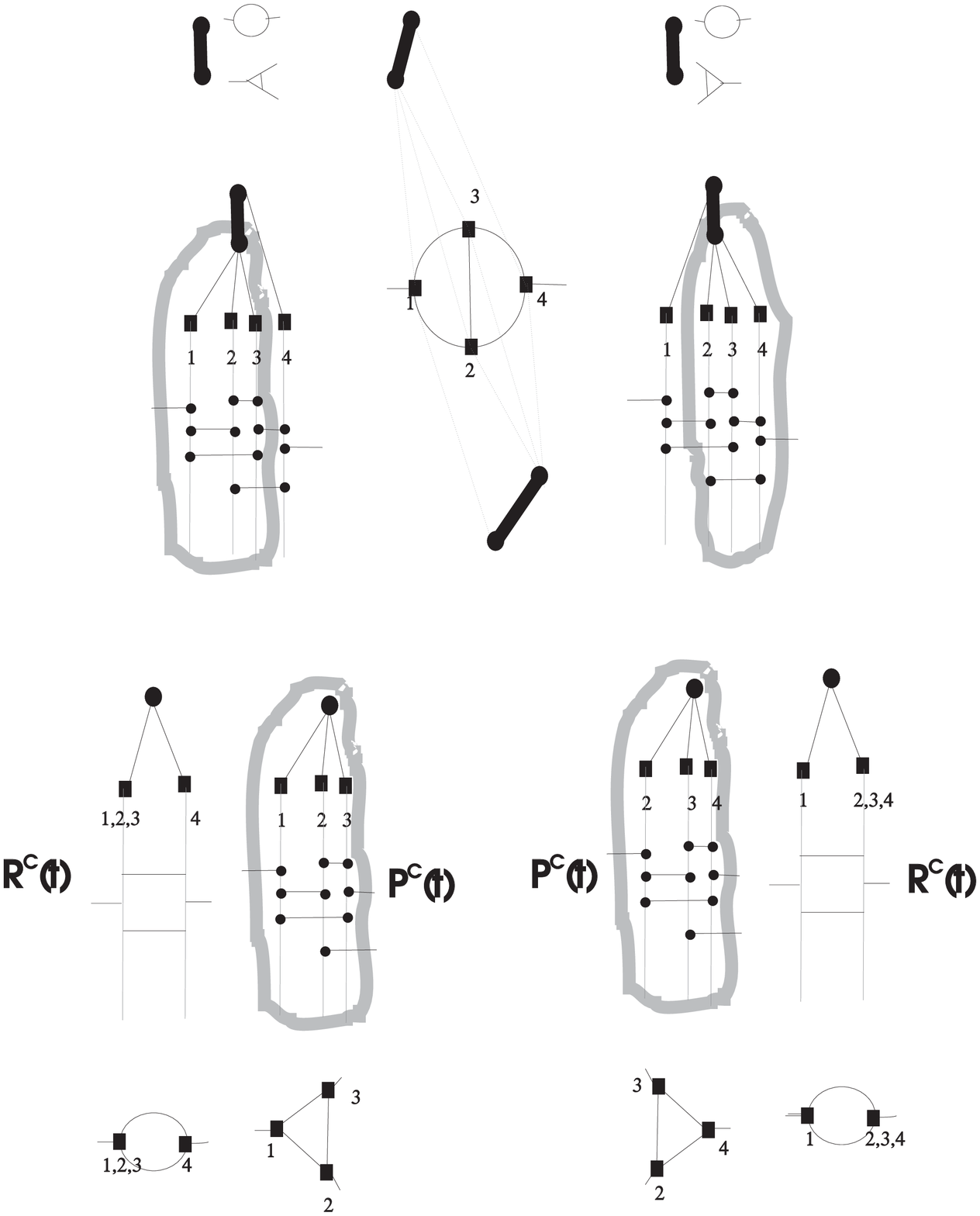}{}{f3}{The graph
$\Omega$ gives rise to two rooted trees corresponding to its two
(overlapping) divergent sectors. Each of the two rooted trees
allows for a single admissible cut. We implement it in each case
by the gray curve which encircles the one vertex which constitutes
$P^C(t)$ and the whole chord diagram attached to this subtree. It
hence corresponds to a subgraph which is a three-point graph, as
three chords are crossed by this gray curve. The cut at the rooted
tree then corresponds to shrinking the subgraph to a point, which
is a vertex in the remaining graph (a one-loop self-energy). This
vertex we have decorated by $\{2,3,4\}$ or $\{1,2,3\}$. It amounts
to a local polynomial insertion in the self-energy. If the
vertices so-generated always give rise to polynomial insertions
which are part of the action already, we have a renormalizable
theory. For a general theory, one will have a variety of different
chords represented by different propagators, and a variety of
vertices as well. For a renormalizable theory there will be only a
finite number of each. It may happen that there are various
different vertices into which a graph can shrink, in which case a
sum over the corresponding external structures is involved
\cite{CK2}.}{12}

Each short-distance singularity corresponds to a sector which can
be described by a rooted tree, which itself notates the hierarchy
of singularities. We have a coproduct which describes the job-list
\cite{BK1} of renormalization: we use it to disentangle the
singularities located at (sub-)diagonals. The Feynman rules are
then providing a character $\phi: H\to V$ on this Hopf algebra.
They map a Hopf algebra element to an analytic expression,
typically evaluating in a suitable ring $V$ of Feynman integrands
or Laurent polynomials in a regularization parameter. These maps
being characters, we have
\be\phi(\gamma_1\gamma_2)=\phi(\gamma_1)\phi(\gamma_2).\ee

Then, renormalization comes from the very simple Hopf algebra
property Eq.(\ref{basic}), as we now explain. Let us describe
carefully how to use the Hopf algebra structure in the example of
Fig.(\ref{f3}). The first thing which we have to introduce,
together with our Feynman rules, is a map $R:V\to V$ which is
essentially determined by the choice of a renormalization scheme.
The freedom in this choice is essentially what makes up the
renormalization group.

The presence of the antipode $S$ allows to consider, for each
$\phi$, its inverse character $\phi^{-1}=\phi\circ S$. Actually,
we have a group structure on characters: to each two characters
$\phi,\psi$ we can assign a new character $$\phi\star\psi=m_V\circ
(\phi\otimes\psi)\circ\Delta,$$ and a unit of the $\star$-product
is provided  as $$ \phi\star \eta=\eta\star\phi=\phi$$ and the
inverse is indeed provided by the antipode:
$$\phi^{-1}\star\phi=\phi\star\phi^{-1}=\eta,$$ where $\eta$ comes
from the counit and is uniquely defined as $\eta=\psi\circ
E\circ\bar{e}$ so that $\eta(1)=1_V$, $\eta(X)=0$, $\forall
X\not=1$, and for any arbitrarily chosen character $\psi$ (any
character fulfills $\psi(1)=1_V$).

The next thing to do is to use $\phi$ and $R$ to define a further
character $S_R:H\to V$ by $$S_R=-R[\phi(t)+\sum
S_R(t^\prime)\phi(t^{\prime\prime})],$$ where we used the notation
$\Delta(t)=t\otimes 1 +1\otimes t+\sum t^\prime\otimes
t^{\prime\prime}$. By construction, if we choose $R={\rm id}_V$,
the identity map from $V\to V$, we have $S_{{\rm id}_V}=\phi\circ
S$.

Now, consider $S_R\star \phi$. We have $$ S_{{\rm
id}_V}\star\phi=\phi\circ m\circ (S\otimes id)\circ\Delta=\eta$$
by the Hopf algebra property Eq.(\ref{basic}) above. This
guarantees that from regions where $R$ becomes the identity map
${\rm id}_V:V\to V$, we get a vanishing contribution from any
non-trivial sector $t$ realized in a Feynman graph $\Gamma$, as
$\eta(t)=0$. So if we demand that $R$ leaves short distance
singularities unaltered, so that $R={\rm id}_V$ for large loop
momenta, we automatically have a vanishing contribution of those
singularities to $S_R\star\phi$.\footnote{That $R$ leaves
short-distance singularities unaltered typically requires that the
first few Taylor coefficients in the Feynman integrands, as
determined by powercounting, are left unaltered.}

What we see at work here is a general principle of multiplicative
subtraction \cite{CK1}: while for a primitive Hopf algebra element
$t$, $\Delta(t)=t\otimes 1+1\otimes t$, $S_R\star\phi$ amounts
simply to the additive operation $$ \phi(t)-R[\phi(t)],$$ for a
general Hopf algebra element the coproduct provides a much more
refined multiplicative subtraction mechanism, which can obviously
be considered for a wide class of Hopf algebras. This principle
can certainly be applied in the future not only in the problem of
short distance singularities, but in a much wider class of
problems, with asymptotic expansions coming to mind immediately.

Fig.(\ref{fn}) describes how the Hopf algebra is realized on the
sectors of the graph $\Omega$ and how this relates to the Hopf
algebra of Feynman graphs to which we now turn.
\bookfig{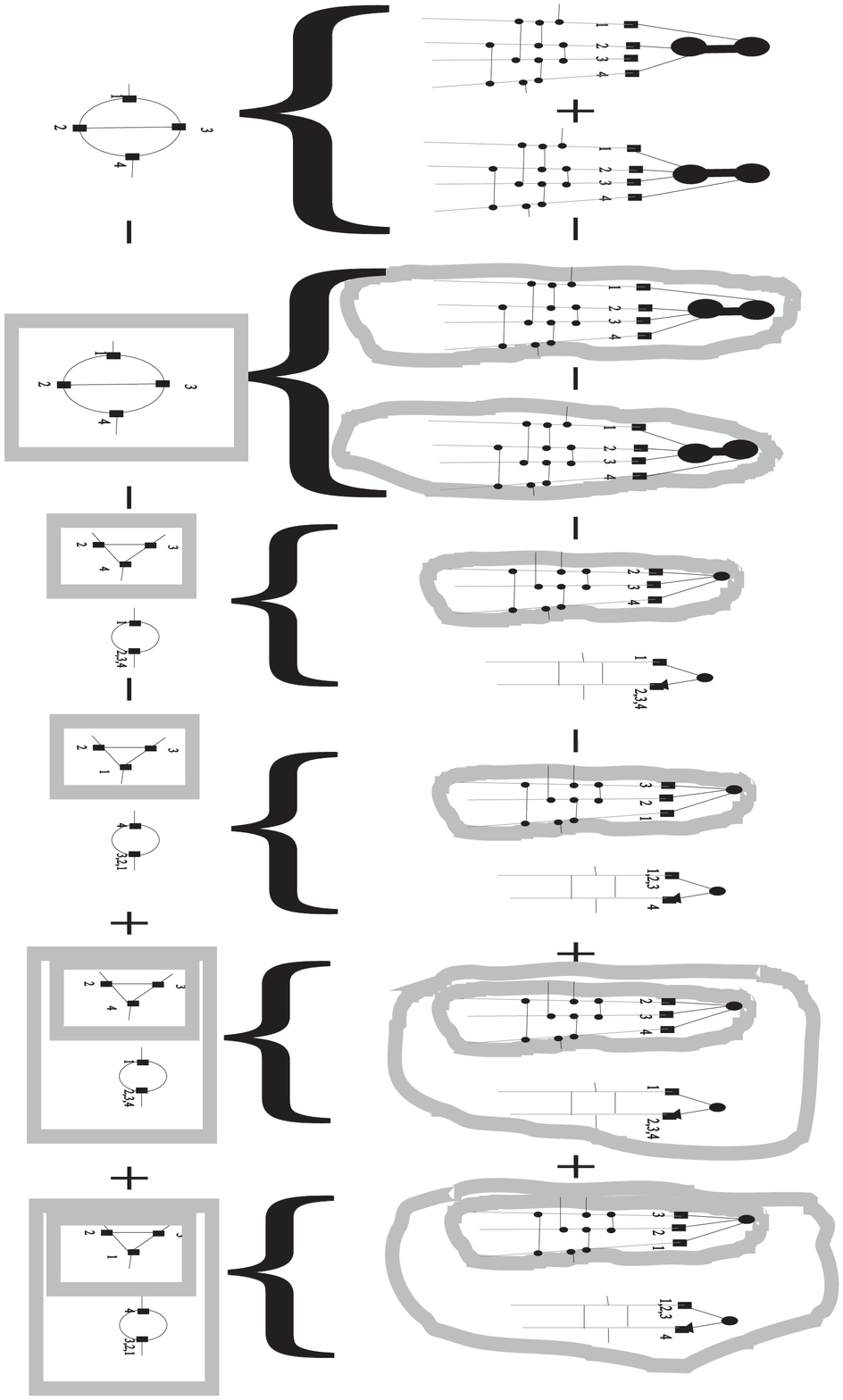}{}{fn}{The result of the operation
$S_R\star\phi(\Omega)$ graphically, where an application of the
operation $R$ is indicated by encircling the graph whose
corresponding analytic expression is to be mapped to the range of
$R$ by a thick grey line. In the upper row, we see the result in
terms of the decorated rooted trees of Fig.(\ref{f3}) while in the
second row we see the result directly expressed in terms of
Feynman integrals. Again, the map $\phi$ is not explicitly written
out. The grey boxes indicate the full and normal forests of
classical renormalization theory \cite{Collins} and are in
one-to-one correspondence with the cuts at the corresponding
rooted trees if we incorporate the empty and the full cut in the
sum over cuts, so that the two terms $T\otimes 1+1\otimes T$ which
appear in any coproduct $\Delta(T)=T\otimes 1+1\otimes
T+\sum_{adm.C}P^C(T)\otimes R^C(T)$ can be regarded as generated
by the full ($T\otimes 1$) and the empty cut ($1\otimes T$)
\cite{CK1}.}{14}

\subsection{The Hopf algebra of graphs}
As we already have emphasized the Hopf algebra of rooted trees is
the role model for the Hopf algebras of Feynman graphs which
underly the process of renormalization when formulated
perturbatively at the level of Feynman graphs. The following
formulas should be of no surprise after our previous discussions.

First of all, we start considering one-particle irreducible graphs
as the linear generators of the Hopf algebra, with their disjoint
union as product. We then define a Hopf algebra by a coproduct
\be\Delta(\Gamma)=\Gamma\otimes
1+1\otimes\Gamma+\sum_{\gamma{\subset}
\Gamma}\gamma\otimes\Gamma/\gamma,\ee where the sum is over all
unions of one-particle irreducible (1PI) superficially divergent
proper subgraphs and we extend this definition to products of
graphs so that we get a bialgebra \cite{CK2}. The above sum
should, when needed, also run over appropriate external structures
to specify the appropriate type of local insertion \cite{CK2}
which appear in local counterterms, which we omitted in the above
sum for simplicity.\footnote{A simple example exhibited in
\cite{CK2} is the self-energy in massive $\phi^3$ theory in six
dimensions. It provides two external structures, corresponding to
local insertions of counterterms for the $m^2\phi^2$ and for the
$(\partial_\mu\phi)^2$ term.}

The counit $\bar{e}$ vanishes, as before, on any non-trivial Hopf
algebra element. At this stage we have a commutative, but
typically not cocommutative bialgebra. It actually is a Hopf
algebra as the antipode in such circumstances comes almost for
free as \be S(\Gamma)=-\Gamma-\sum_{\gamma{\subset}
\Gamma}S(\gamma)\Gamma/\gamma.\ee The next thing we need are
Feynman rules, which we regard as maps $\phi:H\to V$ from the Hopf
algebra of graphs $H$ into an appropriate space $V$.

Over the years, physicists have invented many calculational
schemes in perturbative quantum field theory, and hence it is of
no surprise that there are many choices for this space. For
example, if we want to work on the level of Feynman integrands in
a BPHZ scheme, we could take as this space a suitable space of
Feynman integrands (realized either in momentum space or
configuration space, whatever suits). An alternative scheme would
be the study of regularized Feynman integrals, for example the use
of dimensional regularization would assign to each graph a
Laurent-series with poles of finite order in a variable $\ve$ near
$\ve=0$, and we would obtain characters evaluating in this ring.
In any case, we will have
$\phi(\Gamma_1\Gamma_2)=\phi(\Gamma_1)\phi(\Gamma_2)$.

Then, with the calculational scheme chosen and the Feynman rules
providing a canonical character $\phi$, we will have to make one
further choice: a renormalization scheme. This is is a map $R:V\to
V$, and we demand that is does not modify the UV-singular
structure: in BPHZ language, it should not modify the Taylor
expansion of the integrand for the first couple of terms divergent
by powercounting. In dimensional regularization, we demand that it
does not modify the pole terms in $\ve$.

Finally, the principle of multiplicative subtraction works as
before: we define a further character $S_R$ which deforms
$\phi\circ S$ slightly and delivers the counterterm for $\Gamma$:
\be S_R(\Gamma)=-R[\phi(\Gamma)]-R\left[\sum_{\gamma{\subset}
\Gamma}S_R(\gamma)\phi(\Gamma/\gamma)\right]\ee which should be
compared with the undeformed \be \phi\circ
S=-\phi(\Gamma)-\sum_{\gamma{\subset} \Gamma}\phi\circ S
(\gamma)\phi(\Gamma/\gamma).\ee Then, the classical results of
renormalization theory follow suit \cite{DK1,DKo,CK1}. We obtain
the renormalization of $\Gamma$ by the application of a
renormalized character $$\Gamma\to S_R\star\phi(\Gamma)$$ and the
$\bar{R}$ operation as \be\bar{R}(\Gamma)=\phi(\Gamma)+
\sum_{\gamma{\subset}\Gamma} S_R(\gamma)\phi(\Gamma/\gamma),\ee so that
we have \be S_R\star\phi(\Gamma)=\bar{R}(\Gamma)+S_R(\Gamma).\ee

In the above, we have given all formulas in their recursive form.
Zimmermann's original forest formula solving this recursion is
obtained when we trace our considerations back to the fact that
the coproduct of rooted trees can be written in non-recursive
form, and similarly the antipode. It is not difficult to see that
the sum over all cuts corresponds to a sum over all forests, and
the notion of full and normal forests of Zimmermann \cite{Collins}
gives rise to appropriate sums over cuts \cite{DK1,CK1}, making
use of the graphical implementation of cuts as for example in
Fig.(\ref{fn}).

\section{Rescalings and renormalization schemes}
Let us come back to unrenormalized Feynman graphs, and their
evaluation by some chosen character $\phi$, and let us also choose
a renormalization scheme $R$. The group structure of such
characters on the Hopf algebra can be used in an obvious manner to
describe the change of renormalization schemes. This has very much
the structure of a generalization of Chen's Lemma \cite{DK2}.
\subsection{Chen's Lemma}
Consider $S_R\star\phi$. Let us change the renormalization scheme
from $R$ to $R^\prime$. How is the renormalized character
$S_{R^\prime}\star\phi$ related to the renormalized character
$S_R\star\phi$? The answer lies in the group structure of
characters: \be S_{R^\prime}\star\phi=[S_{R^\prime}\star S_R\circ
S]\star [S_R\star\phi].\ee We inserted a unit $\eta$ with respect
to the $\star$-product in form of $\eta=S_R\circ S\star S_R\equiv
S_R^{-1}\star S_R$, and can now read the rerenormalization,
switching between the two renormalization schemes, as composition
with the renormalized character $S_{R^\prime}\star
S_R^{-1}$.\footnote{$S_{R^\prime}\star S_R\circ S$ is a
renormalized character indeed: if $R,R^\prime$ are both self-maps
of $V$ which do not alter the short-distance singularities as
discussed before, then in the ratio $S_{R^\prime}\star S_R\circ S$
those singularities drop out.}

Similar considerations apply to a change of scales which determine
a character \cite{DK2}. If $\rho$ is a dimensionful parameter
which appears in a character
$\phi=\phi(\rho)$,\footnote{Typically, it could be a scale which
dominates the process under consideration.} then the transition
$\rho\to\rho^\prime$ is implemented in the  group by acting on the
right with the renormalized character
$\psi_{\rho,\rho^\prime}^\phi:=\phi(\rho)\circ
S\star\phi(\rho^\prime)$ on $\phi(\rho)$,
\be\phi(\rho^\prime)=\phi(\rho)\star
\psi_{\rho,\rho^\prime}^\phi.\ee Let us note that this Hopf
algebra structure can be efficiently automated as an algorithm for
practical calculations  exhibiting the full power of this
combinatorics \cite{BK1}.

Now, assume we compute Feynman graphs by some Feynman rules in a
given theory and decide to subtract UV singularities at a chosen
renormalization point $\mu$. This amounts, in our language, to
saying that the map $S_R$ is parametrized by this renormalization
point: $S_R=S_R(\mu)$. Then, let $\Phi(\mu,\rho)$ be the ratio
$\Phi(\mu,\rho)=S_R(\mu)\star\phi(\rho)$. We then have the
groupoid law generalizing the before-mentioned Chen's lemma
\cite{DK2} \be
\Phi(\mu,\eta)\star\Phi(\eta,\rho)=\Phi(\mu,\rho).\ee While this
looks like a groupoid law, the product of two unrelated ratios
$\Phi(\mu_1,\mu_2)$ $\star\Phi(\mu_3,\mu_4)$, as any other product
of characters, is always well-defined in the group of characters
of the Hopf algebra.
\subsection{Automorphisms of the Hopf algebra}
 In the set-up discussed
so far, the combinatorics of renormalization was attributed to a
Hopf algebra, while characters of this Hopf algebra took care of
the specific Feynman rules and chosen renormalization schemes.
Renormalized quantities appear as the ratio of two characters,
while divergences drop out in this ratio $S_R\star\phi$.

Typically, such characters introduce a renormalization scale
(cut-off, the 't Hooft mass $\mu$ in dimensional regularization),
and we can use these parameters to describe the change of schemes
in a fairly unified manner, as discussed in \cite{DK2}.

These considerations of changes of renormalization schemes are
related to another interesting aspect discussed in \cite{DK2}. So
far, we regarded the map $R$ as a self-map in a certain space $V$.
We will not have $R(XY)=R(X)R(Y)$ (for example, minimal
subtraction cannot possibly fulfill that the poleterms of a
product is the product of the poleterms), but $R$ obeys the
multiplicativity constraints \be
R(XY)+R(X)R(Y)=R(XR(Y))+R(R(X)Y),\ee which ensure that
$S_R(\Gamma_1\Gamma_2)=S_R(\Gamma_1)S_R(\Gamma_2)$ \cite{DK2,CK2}.
This leads to the Riemann--Hilbert problem to be discussed below.

We now want to investigate to what extent the map $R:V\to V$ can
be lifted to an automorphism $\Theta_R:H\to H$ of the Hopf
algebra. We regard $V$ as the space in which Feynman graphs
evaluate by the Feynman rules, as discussed above. Let again the
Feynman rules be implemented by $\phi$. The map $S_R$ is then a
character constructed with the help of $\phi$, so we should write
$S_R\equiv S_R^\phi$ to be exact.

The question is if one can construct, for any $R$, an automorphism
$\Theta_R:H\to H $ of the Hopf algebra such that one has \be
\overline{\phi}\equiv S_R^\phi\circ S=\phi\circ \Theta_R,\ee so
that (using $S^2={\rm id}$ which is true in any commutative Hopf
algebra) \be S_R\star\phi=\overline{\phi}\circ
S\star\phi=\phi\circ\Theta_R\circ
S\star\phi=\phi\circ[\Theta_R^{-1}\star{\rm id}].\ee The answer is
affirmative \cite{DK2}. Following \cite{DK2} and the use of
one-parameter group of automorphisms in the renormalization group
\cite{CK3}  to be discussed below, we make the following {\em
Ansatz} for $\Theta_R$: \be\Theta_R(\Gamma)=\Gamma e^{-\ve{\rm
deg(\Gamma)} \rho_R(\Gamma)},\ee where, in the context of
dimensional regularization or any other analytic regularization,
$\rho_R(\Gamma)$ will be a character evaluating in the ring of
Taylor series in $\ve$ regular at $\ve=0$ and ${\rm
deg}(\Gamma)=n$ if $\Gamma$ has $n$ loops.\footnote{It is
convenient but not necessary to work with dimensional
regularization here. In BPHZ, one could work for example with the
ratio of Taylor series in external parameters.} Then, one
determines \be\rho_R(\Gamma)=\frac{-1}{\ve{\rm deg}(\Gamma)
}\log\left(\frac{S_R^\phi\circ S(\Gamma)}{\phi(\Gamma)}\right),\ee
so that indeed $\rho_R(\Gamma)$ is free of poleterms, as one
easily shows $$\frac{S_R^\phi\circ
S(\Gamma)}{\phi(\Gamma)}=1+{\cal O}(\ve),$$ for arbitrary graphs
$\Gamma$. This gives a unifying approach to the treatment of
renormalization schemes and changes between them.\footnote{From
here, one can start considering categorical aspects of
renormalization theory and in particular address the question
posed in \cite{DK1} if a modified coproduct
$\Delta_R=(\Theta_R\otimes{\rm id})\circ\Delta$ is
(weak-)coassociative in dependence of $R$, with first results
upcoming  in a recent thesis \cite{Mertens}.}

\section{The insertion operad of Feynman graphs}
In this section, we want to describe an operad structure on
Feynman graphs. This operad was implicitly present in many results
in \cite{CK1,CK2,CK3}, and so it is worth to describe it shortly
at this stage, also with regard to the fact that it will prove to
be a useful construct to investigate the number-theoretic aspects
of Feynman graphs \cite{DK3,DK4} to be discussed below.

While the previous two sections discussed the process of
disentangling a Feynman graph into subgraphs according to the
presence of UV singularities, we now turn to the process of
plugging graphs into each other. This will lead us in the next
section to Lie algebras of Feynman graphs. Here, we want to study
the most basic operation: plugging one graph $\Gamma_1$ into
another graph $\Gamma_2$. Typically, there are various places in
$\Gamma_2$,  provided by edges and vertices of $\Gamma_2$, which
can be replaced by $\Gamma_1$. To obtain a sensible notion of this
operation we should fulfill operad laws in this process. These
operad laws can be described as follows. Operad laws are concerned
with rules which should be fulfilled when we insert several times.
First, assume we have graphs $\gamma_1,\gamma_2$ and want to plug
both of them into different places of a graph $\Gamma$. Then, the
result should be independent of the order in which we do it. Next,
when we plug $\gamma_1$ into $\gamma_2$ at some place, and insert
the result into $\Gamma$, the result should be the same as
inserting $\gamma_2$ at the same place in $\Gamma$, and then
$\gamma_1$ into the corresponding relabelled place of $\gamma_2$.
Finally, the permutation of places should be compatible with the
composition (see for example \cite{L} for a formal definition of
these requirements).

We only describe the operad in the context of massless $\phi^3$
theory in six dimensions, the generalizations to more general
cases are obvious and will be discussed elsewhere.

A Feynman graph provides vertices and edges connecting these
vertices. The operad  essentially consists of regarding these
vertices and edges as places into which other graphs can be
inserted. Naturally, a vertex correction can replace a vertex of a
similar type, and a propagator-function can replace a line which
represents a free propagator of a similar type. In massless
$\phi^3$ theory, we only have one type of lines and one type of
vertices.

First, we note that the overall divergent Feynman graphs in this
theory are given by 1PI graphs with two or three amputated
external lines. Thus, vertices in the graphs are either internal
three-point vertices, or two-point vertices resulting from the
amputation of an external leg from a three-point vertex.
 Hence, self-energies can be described as graphs which precisely have
two two-point vertices, while three-point graphs, --vertex
corrections--, have precisely three two-point vertices.
Propagator-functions then have two external edges.

When we want to replace an internal vertex, we just replace it by
a vertex correction. When we want to replace an internal edge, a
free propagator, we replace it by a propagator-function, as
described by Figs.(\ref{f4},\ref{f5}).

How many places are there? Let $\Gamma(p_1,p_2)$ be  a 1PI vertex
function given by a three point graph $\Gamma$ with $l$ loops,
which then provides $2l+1$ vertices and $3l$ internal lines, hence
$5l+1$ places for insertion altogether. Let $\Pi(p)$ be a
propagator function given by a (not necessarily one-particle
irreducible) two-point graph $\Pi$ with $l$ loops, it then
provides $2l$ vertices and $3l+1$ lines, hence again $5l+1$ places
(we not necessarily have to label all edges and vertices, for
example dropping the label at an external edge of the propagator
function takes into account quite naturally the fact that
self-energies are proportional to an inverse propagator, and, in a
massless theory, cancel one of the external lines).

We label all edges and vertices in arbitrary order, and the
composition laws described in the figure captions of
Figs.(\ref{f4},\ref{f5}) fulfill the operad laws (the
before-mentioned requirements  are fulfilled), so that
Figs.(\ref{f4},\ref{f5}) define this operad by way of example.
\bookfig{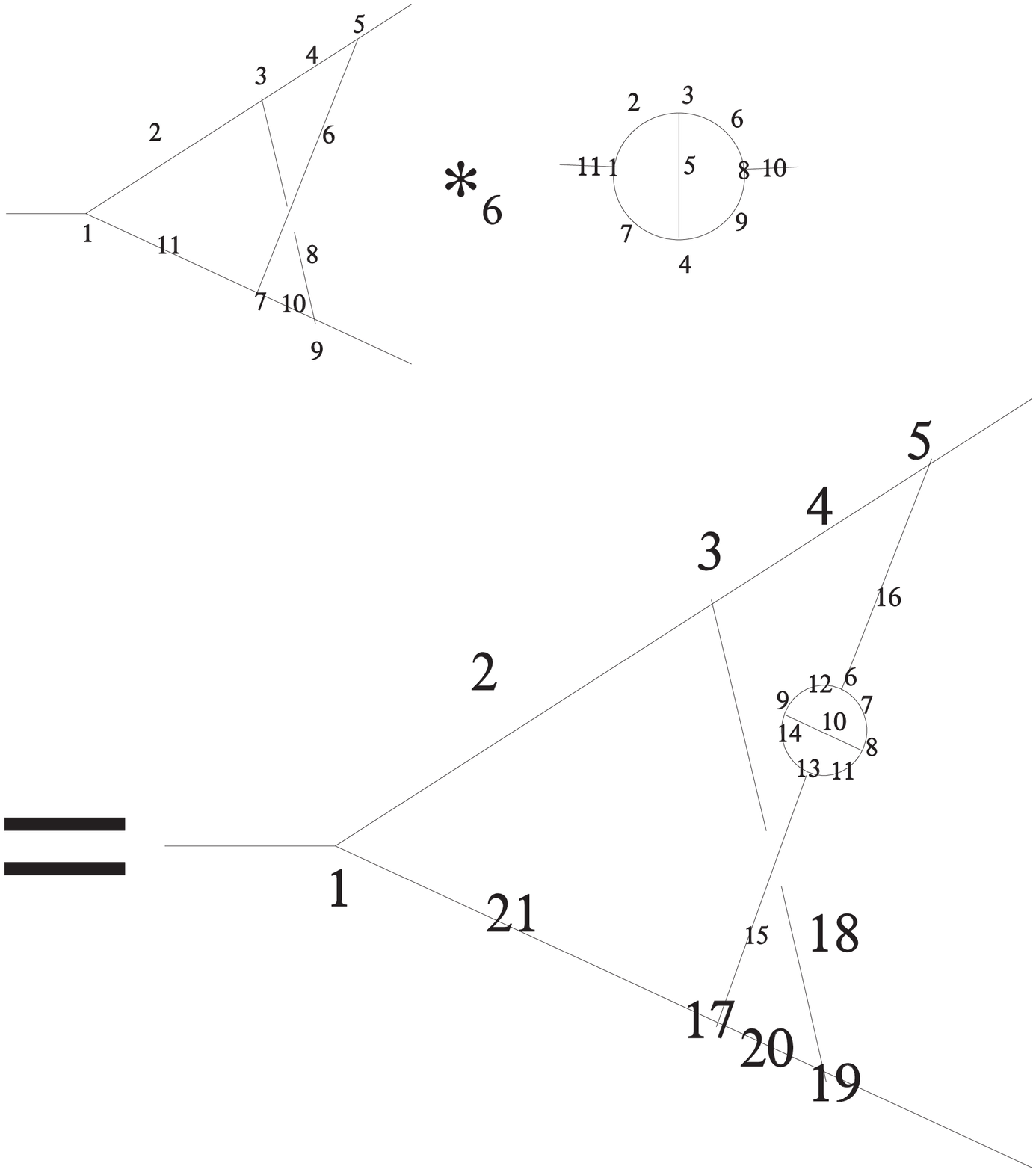}{}{f4}{We consider a propagator-graph $\gamma$
and a vertex-function $\Gamma$ and as an example  their
concatenation $\Gamma\star_6\gamma$. The propagator function
replaces the line with label 6 in the vertex-function. The
propagator-function provides four vertices (labelled 1,3,4,8) and
seven edges (labelled 2,5,6,7,9,10,11). Two of the edges, 10 and
11, are external. The vertex-function provides five vertices
(labelled 1,3,5,7,9) and six edges (labelled 2,4,6,8,10,11). The
vertices 1,5,9 are external, they connect to edges which are not
part of the vertex function. We still indicated them by open-ended
lines at those vertices, but one should regard vertices 1,5,9 as
two-point vertices.

Note that each internal edge ends in two labelled vertices. We
replace the edge labelled 6 by the propagator-function, connecting
the external edges 10 and 11 of the latter to the vertices 5 and 7
of the vertex-function. We glue the edge with the lower label (10)
to the vertex with the lower label (5). Relabelling is done in the
obvious way: labels 1 to 5 in the vertex-function remain
unchanged, the labels at the inserted propagator function become
labels 6 to 16, and labels 7 to 11 become labels 17 to 21,
increasing their labels by $4+7-1=10$.}{11.2}
\bookfig{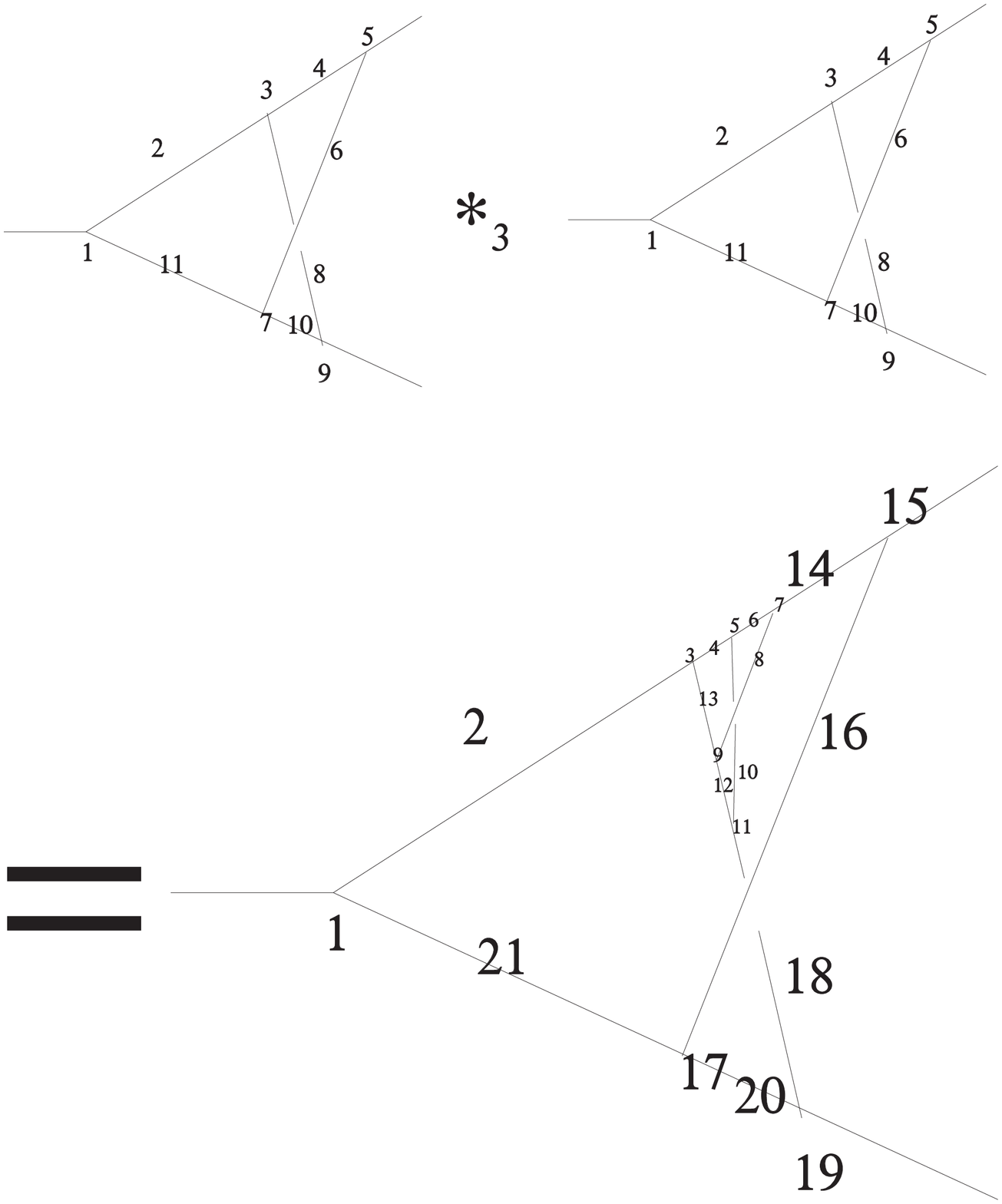}{}{f5}{To explain the insertion of a
vertex-function, we replace in this example vertex 3 of a
vertex-function by the very same vertex-function, so we describe
$\Gamma\star_3\Gamma$. We do it by connecting edges 2,4,8 which
are attached to vertex 3 to the three two-point vertices 1,5,9,
respecting the order: edge 2 connects to vertex 1, edge 4 to
vertex 5, edge 8 to vertex 9. Relabelling is done in the obvious
way: labels 1 and 2 in the vertex-function remain unchanged, the
labels at the inserted vertex-function become labels 3 to 13, and
labels 4 to 11 become labels 14 to 21.}{12.5}

So, with these rules for insertion (we also understand that
insertion of a propaga\-tor-function at a vertex place or a
vertex-function at an edge vanishes trivially by definition), one
gets indeed an (partial) operad. Note further that insertion of a
free propagator or vertex leaves the result unchanged.

One easily extends this construction to the case that one has
vertices of other valencies and with different sorts of lines
coming in.

This operad can be conveniently used to study the Lie algebraic
structure of diagrams as well as for the investigation of
number-theoretic aspects as we will see below. Also, the operad
viewpoint is helpful in understanding the equivalence classes
discussed in \cite{DK1}. For example, the two graphs $\Gamma_1$
and $\Gamma_2$ of section 2.1 belong to the same equivalence
class, $\Gamma_1\sim\Gamma_2$,  given by the parenthesized word
$((\gamma)\Gamma_0)$, and are distinguished only by the place into
which we insert $\gamma$. In general, two graphs are equivalent if
one is obtained via a permutation of concatenation labels of the
other, while maintaining the tree structure of its subdivergences:
all Feynman graphs which represent the same rooted tree or
parenthesized word can be obtained from each other by the change
of labels of places where we insert the primitive graphs into each
other.

Also, typical equations in field theory like Schwinger-Dyson
equations are naturally formulated by this operad, using the fact
that the sum over all diagrams can be written as a sum over all
primitive ones into which all diagrams are plugged in all possible
places. Details will be given in future work.

\section{The Lie algebra structure}
In \cite{CK1,CK2,CK3} the reader finds various Lie algebra
structures which appear in the dual of the Hopf algebra which is
the universal enveloping algebra of a Lie algebra. Here, we
describe the Lie algebra of Feynman graphs. There is also one for
rooted trees, which can be found in \cite{CK1}.

Study of these Lie algebras is a very convenient way of
understanding the structure of Feynman graphs. These Lie algebras
play a crucial role when one wants to understand the connection
between the group of diffeomorphisms of physical parameters like
coupling constants with the group of characters of the Hopf
algebra, to which we will turn in the next section.

It is also quite useful in determining the Hopf algebra structure
of a chosen QFT correctly, because, once it is found, the
corresponding enveloping algebra will be the dual of a commutative
non-cocommutative Hopf algebra (by the celebrated Milnor--Moore
theorem \cite{CK1,CK2}) whose coproduct gives us the forests
formulas of renormalization.\footnote{For example one easily
determines the Lie algebra of QED, having one type of vertex
connecting to two different type of lines for fermion and photon
propagators. This then confirms the corresponding Hopf algebra
structure of 1PI graphs to be commutative non-cocommutative.
One-particle reducible graphs can be treated as in \cite{KD}. In
the literature, there are other attempts to describe the
renormalization of QED by binary rooted trees \cite{Brouqed}. But
the singularities of QED are stratified along diagonals as in any
local QFT, and  the rather artificial restriction to binary rooted
trees ultimately runs into trouble \cite{CBcomm}.} To find these
Lie algebras, one defines a Lie-bracket of two 1PI graphs
$\Gamma_1,\Gamma_2$ by plugging $\Gamma_1$ into $\Gamma_2$ in all
possible ways and subtracts all ways of plugging $\Gamma_2$ into
$\Gamma_1$.

These Lie algebras all arise from a pre-Lie structure which we can
describe in Fig.[\ref{f6}]. The operation of inserting one graph
$\Gamma_1$ in another graph $\Gamma_2$ in all possible ways is a
pre-Lie operation $\Gamma_2\star\Gamma_1$ , which means that it
fulfills $$\Gamma_3\star (\Gamma_2\star\Gamma_1)-(\Gamma_3\star
\Gamma_2)\star\Gamma_1 =\Gamma_3\star
(\Gamma_1\star\Gamma_2)-(\Gamma_3\star \Gamma_1)\star\Gamma_2.$$
Antisymmetrization then gives automatically a bracket
$[\Gamma_1,\Gamma_2]=\Gamma_1\star\Gamma_2-\Gamma_2\star\Gamma_1$,
which fulfills the Jacobi identity. This operation of inserting
one graph in another in all possible ways can obviously written
with the help of the operad structure of the previous section as a
sum over all places where to insert (plus a sum over all
permutations of the labels of identical external vertices of the
graph which is to be inserted) and the operad laws then guarantee
that the pre-Lie property is fulfilled, making use of the intimate
connection between rooted trees, operads and pre-Lie algebras
\cite{CL}.

Once this Lie algebra is found, one knows that dually one obtains
a commutative, non-cocommutative Hopf algebra which is the basis
of the forest formulas of renormalization as discussed in the
previous section.  \bookfig{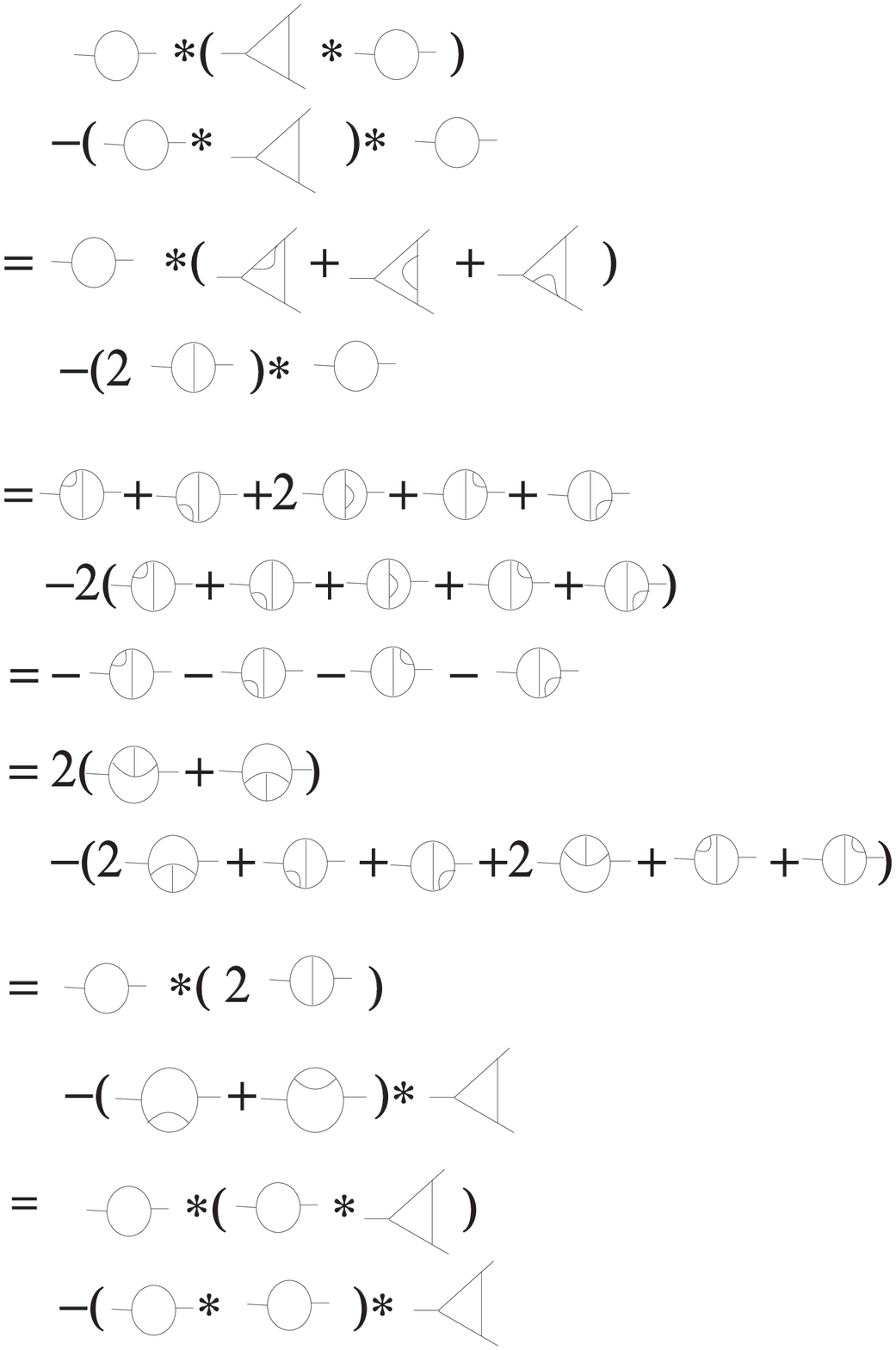}{}{f6}{The (pre-)Lie algebra
structure of Feynman graphs. The fact that the operation of
plugging a graph into another one in all possible ways is pre-Lie
is essentially due to the fact that the ways of plugging (in all
possible ways) $\Gamma_1$ into $\Gamma_2$, and the result into
$\Gamma_3$, subtracted from the ways of plugging (in all possibly
ways) $\Gamma_1$ into the result of plugging (in all possible
ways) $\gamma_2$ into $\Gamma_3$ is the sum over all possible ways
to plug $\Gamma_1,\Gamma_2$ disjointly into $\Gamma_3$.}{13}

It is not difficult to work out the corresponding pre-Lie
structure for QED for example, and indeed, reading the graphs of
Fig.(\ref{f6}) as QED graphs in the obvious possible manners only
demands to cancel a few of the terms in that figure, because a
photon propagator can only replace a photon line, and not a
fermion line. Similarly, for any local QFT, one can determine the
corresponding Hopf and Lie algebras, incorporating external
structures whenever necessary as in \cite{CK2}.

The resulting Lie algebras of Feynman graphs play a fundamental
role in understanding how the combinatorial properties of
renormalization connect to the renormalization group, to the
running of physical parameters. We now turn to study these results
of \cite{CKRH,CK2,CK3}.
\section{The Birkhoff decomposition and the renormalization group}
In \cite{CKRH,CK2,CK3} the reader finds an amazing connection
between the Riemann--Hilbert problem and renormalization. This
result was first announced in \cite{CKRH}. It is based on the use
of a complex regularization parameter. Typically, dimensional
regularization provides such a parameter as the deviation $\ve$
from the relevant integer dimension of spacetime, but for example
analytic regularization would do as well.

With such a regularization parameter, the Feyman rules map a
Feynman graph to a Laurent series with poles of finite order in
this regularization parameter, hence the Feynman rules provide a
character from the Hopf algebra of Feynman graphs to the ring of
Laurent polynomials with poles of finite order in $\ve$.

As mentioned before, the multiplicativity constraints \cite{DK2,CKRH,CK2} $$
R[xy]+R[x]R[y]=R[R[x]y]+R[xR[y]]$$ ensure that the corresponding
counterterm map $S_R$ is a character as well,
$$S_R[xy]=S_R[x]S_R[y],\;\forall x,y\in H.$$ We now study how this
set-up leads to the Riemann--Hilbert problem and the Birkhoff
decomposition.
\subsection{Minimal subtraction: the Birkhoff decomposition}
To make contact with the Riemann--Hilbert problem, the crucial
step is to recognize that, for $R=MS$ being chosen to be
projection onto these poles of finite order (the minimal
subtraction scheme MS), $\phi=S_{MS}\circ S\star[S_{MS}\star\phi]$
is a decomposition of the character $\phi$ into a part which is
holomorphic at $\ve=0$: $S_{MS}\star\phi\equiv\phi_+$ is a
character evaluating in the ring of functions holomorphic at
$\ve=0$, while $S_{MS}\equiv\phi_-$ maps to polynomials in $1/\ve$
without constant term, it delivers, when evaluated on Feynman
graphs, the MS counterterms for those graphs. This corresponds to
a Birkhoff decomposition $\phi=\phi_-^{-1}\phi_+$. For an
introduction to the Riemann--Hilbert problem and the associated
Birkhoff decomposition we refer the reader to \cite{Boli}.
Suffices it here to say that the Riemann--Hilbert problem is a
type of inverse problem. For a given complex differential equation
$$y^\prime(z)=A(z)y(z),\;A(z)=\sum_i\frac{A_i}{z-z_i}$$ with given
regular singularities $z_i$ and matrices $A_i$, one can determine
monodromy matrices $M_i$ integrating around curves encircling the
singularities. The inverse problem, finding the differential
equation from knowledge of the singular places and monodromy
matrices, is the Riemann--Hilbert problem. A crucial role in its
solution plays the Birkhoff decomposition: for a closed curve $C$
in the Riemann sphere, and a matrix-valued loop $\gamma:z\to
\gamma(z)$ well-defined on $C$, decompose it into parts
$\gamma_{\pm}$ well-defined in the interior/exterior of $C$.

Thus, renormalization in the MS scheme can be summarized in one
sentence: with the character $\phi$ given by the Feynman rules in
a suitable regularization scheme and well-defined on any small
curve around $\ve=0$, find the Birkhoff decomposition
$\phi_+(\ve)=\phi_-\phi$, where now and in the following the
product in expressions like  $\phi_-\phi$ is meant to be just the
convolution product $\phi_-\star\phi$ of characters used before.

The unrenormalized analytic expression for a graph $\Gamma$ is
then $\phi[\Gamma](\ve)$, the MS-counterterm is $S_{{
MS}}(\Gamma)\equiv\phi_-[\Gamma](\ve)$ and the renormalized
expression is the evaluation  $\phi_+[\Gamma](0)$. Once more, note
that the whole Hopf algebra structure of Feynman graphs is present
in this group: the group law demands the application of the
coproduct, $\phi_+=\phi_-\phi\equiv S_{MS}\star\phi$.

The transition from here to other re\-nor\-ma\-li\-za\-tion
schemes can be achieved in the group of characters in accordance
with our previous considerations in section 3.

But still, one might wonder what a huge group this group of
characters really is. What one confronts in QFT is the group of
diffeomorphisms of physical parameter: low and behold, changes of
scales and renormalization schemes are just such (formal)
diffeomorphisms. So, for the case of a massless theory with one
coupling constant $g$, for example, this just boils down to formal
diffeomorphisms of the form $$g\to \psi(g)=g+c_2 g^2+\ldots.$$ The
group of one-dimensional diffeomorphisms of this form looks much
more manageable than the group of characters of the Hopf algebras
of Feynman graphs of this theory.

Thus, it would be very nice if the whole Birkhoff decomposition
could be obtained at the level of diffeomorphisms of the coupling
constants, and this is what was achieved in \cite{CK3}.

\subsection{The $\beta$-function}
Following \cite{CK2} in the above we have seen that perturbative
renormalization is a special case of a general mathematical
procedure of extraction of finite values based on the
Riemann-Hilbert problem. The characters of the Hopf algebra of
Feynman graphs form a group whose concatenation, unit and inverse
are given by the coproduct, the counit and the antipode. So we can
associate to any given renormalizable quantum field theory an
(infinite dimensional) complex Lie group $G$ of characters of its
Hopf algebra $H$ of Feynman graphs. Passing from the
unrenormalized theory to the renormalized one corresponds to the
replacement of the loop $\ve \ra \g (\ve) \in G$ (obtained by
restricting the character $\phi$ to an arbitrarily chosen curve
$C$ around $\ve=0$) of elements of $G$ obtained from dimensional
regularization (still, $\ve \ne 0$ is the deviation from the
integer dimension of space-time) by the value $\g_+ (\ve)$ of its
Birkhoff decomposition, $\g (\ve) = \g_- (\ve)^{-1} \, \g_+
(\ve)$.

In \cite{CK3} it was shown how to use the very concepts of a Hopf
and Lie algebra of graphs  to lift the usual concepts of the
$\b$-function and renormalization group from the space of coupling
constants of the theory to the complex Lie group $G$. We now
exhibit these results.

The original loop $\ve \ra \g (\ve)$ not only depends upon the
parameters of the theory but also on the additional {\em unit of
mass} $\mu$, --the 't Hooft mass in dimensional regularization--,
required by dimensional analysis.

But although the loop $\g (\ve)$ does depend on the additional
parameter $\mu$, $$ \mu \ra \g (\ve;\mu) \, , $$ the negative part
$\g_{\mu^-}$ in the Birkhoff decomposition, the character
delivering the MS counterterms, $$ \g (\ve;\mu) = \g_{-}
(\ve;\mu)^{-1} \, \g_{+} (\ve;\mu) $$ is actually independent of
$\mu$, \be \frac{\partial}{\partial \mu} \, \g_{-} (\ve;\mu) = 0
\, . \ee This is a remnant of the fact that our Hopf algebra is
constructed so as to achieve local counterterms: $\phi$ is a
character which can be easily shown to be a series in
$\log(q^2/\mu^2)$ so that a remaining $\mu^2$ dependence in MS
counterterms would be accompanied by a remaining $q^2$ dependence,
and would hence violate locality.\footnote{A similar argument
applies when the Feynman rules provide a character parametrized by
several scales. Again, by a group action which is a finite
rerenormalization, we can reduce the unrenormalized theory to a
dependence on a single scale. This reduction can constrain the
renormalization group flow to a submanifold though, in which case
an explicit group action is needed to switch from mass-independent
to mass-dependent renormalization group functions, as it is
well-known \cite{Kraus}.}

The Lie group $G$ turns out to be graded, with grading, $$ \t_\rho
\in {\rm Aut} \, G \ , \quad \rho \in \Rb \, ,  $$ inherited from
the grading of the Hopf algebra $H$ of Feynman graphs given by the
loop number, \be {\rm deg}(\G) = \hbox{loop number of} \ \G \ee
for any 1PI graph $\G$, so that $\t_\rho(\G)=e^{\rho {\rm deg}
(\G)}\G$.\footnote{Here $\rho$ is to be regarded as a constant. If
we promote it to a character evaluating in the ring of functions
holomorphic at $\ve=0$ we obtain the automorphisms used in section
3 to lift the renormalization map $R$ to automorphisms of the Hopf
algebra. Note that a constant $\rho$ is sufficient to describe
momentum schemes for example, using that one only has to use
$\rho=\ve \log(\mu^2/q^2)$ to compensate for the canonical
$q^2$-dependence \cite{DK1,DK2,KD,BK1}.}

This leads to $$ \g (\varepsilon; e^\rho \mu) = \t_{\rho \ve} (\g
(\ve;\mu)) \qquad \fl \, \rho \in \Rb \, , $$ so that the loops
$\g(\mu)$ associated to the unrenormalized theory have the
property that the negative part of their Birkhoff decomposition is
unaltered by the operation, $$ \g (\ve) \ra \t_{\rho\ve} (\g
(\ve)) \, : $$ if we replace $\g (\ve) $ by $\t_{\rho\ve} (\g
(\ve))$ we do not change the negative part of its Birkhoff
decomposition.  A complete characterization of the loops $\g (\ve)
\in G$ fulfilling this invariance can be found in \cite{CK3}. This
characterization only involves the negative part $\g_- (\ve)$ of
their Birkhoff decomposition which by hypothesis fulfills, \be
\g_- (\ve) \, \t_{\rho \ve} (\g_- (\ve)^{-1}) \ \hbox{is
convergent for} \ \ve \ra 0 \, .\ee It is then easy to see that
this defines in the limit $\ve \ra 0$ a one parameter subgroup,
\be F_\rho \in G \, , \ \rho \in \Rb.  \ee Now, the role of the
$\beta$-function is revealed:  the generator $\b := \left(
\frac{\partial}{\partial \rho} \, F_\rho \right)_{\rho=0}$ of this
one parameter group is related to the {\it residue} of the loop
$\g$ \be \build{\rm Res}_{\ve = 0}^{} \g = - \left(
\frac{\partial}{\partial u} \, \g_- \left( \frac{1}{u} \right)
\right)_{u=0}  \ee by the simple equation, \be \b = Y \, {\rm Res}
\, \g \, ,  \ee where $Y = \left( \frac{\partial}{\partial \rho}
\, \t_\rho \right)_{\rho=0}$ is the grading. In a moment, we will
see how this generator $\beta$ relates to the common
$\beta$-function of physics.

All this is a simple consequence of the set-up described so far
and is worked out in detail in \cite{CK3} (essentially, at the
moment we quote a summary of the results of that paper), while the
central result of \cite{CK3} gives $\g_- (\ve)$ in closed form as
a function of $\b$. Let us use an additional generator in the Lie
algebra of $G$ (i.e.~primitive elements of $H^*$) implementing the
grading such that $ [Z_0 , X] = Y(X) \fl \, X \in \hbox{Lie} \ G
.$ Then, the loop $\g_- (\ve)$ corresponding to the MS counterterm
evaluated on any close curve around $\ve=0$ can be written by a
scattering type formula for $\g_- (\ve)$ as \be \g_- (\ve) =
\lim_{t \ra \ify} e^{-t \left( \frac{\b}{\ve} + Z_0 \right)} \,
e^{t Z_0} \, .  \label{scatt}\ee Both factors in the right hand
side belong to the semi-direct product, $$ \wt G = G \, \semi_{\t}
\, \Rb  $$ of the group $G$ by the grading, but their product
belongs to the group $G$.

As a consequence the higher pole structure of the divergences is
uniquely determined by the residue and this gives a strong form of
the t'Hooft relations, which come indeed as an immediate
corollary.\footnote{The explicit formulas in \cite{CK3} allow to
find the combinations of primitive graphs into which higher order
poles resolve. The weights are essentially given by iterated
integrals which produce coefficients which generalize the
tree-factorials obtained for the undecorated Hopf algebra in
\cite{DK2,KD,BK1}. Iterated application of this formula allows to
express inversely the first-order poles contributing to the
$\beta$-function as polynomials in Feynman graphs free of
higher-order poles.}

The most fundamental result of \cite{CK3} is obtained though when
considering two competing Hopf algebra structures: diffeomorphisms
of physical parameters carry, being formal diffeomorphisms, with
them the Hopf algebra structure of such diffeomorphisms. This
structure was recognized for the first time by Alain Connes and
Henri Moscovici in \cite{CM}. On the other hand, a variation of
physical parameters induced by a variation of scales is a
rerenormalization, which directly leads to the Hopf algebra of
Feynman graphs. Let us first describe the Hopf algebra structure
of the composition of diffeomorphisms in a fairly elementary way,
while mathematical detail can be found in \cite{CM}.

Assume you have formal diffeomorphisms $\phi,\psi$ in a single
variable \be x\to\phi(x)=x+\sum_{k>1}c_k^\phi x^k,\label{expa}\ee
and similarly for $\psi$. How do you compute the Taylor
coefficients $c^{\phi\circ\psi}_k$ for the composition
$\phi\circ\psi$ from the knowledge of the Taylor coefficients
$c_k^\phi,c_k^\psi$? It turns out that it is best to consider the
Taylor coefficients
\be\delta_k^\phi=\log(\phi^\prime(x))^{(k)}(0)\label{CMT}\ee
instead, which are as good to recover $\phi$ as the usual Taylor
coefficients. The answer lies then in a Hopf algebra structure:
$$\delta^{\phi\circ\psi}_k=m\circ(\tilde{ \psi}\otimes\tilde{\phi}
)\circ\Delta_{CM}(\delta_k), $$ where $\tilde{\phi},\tilde{\psi}$
are characters on a certain Hopf algebra $H_{CM}$ (with coproduct
$\Delta_{CM}$) so that $\tilde{\phi}(\prod_i \delta_i)$ $=$
$\prod_i\delta_i^\phi$. Thus one finds a Hopf algebra with
abstract generators $\delta_n$ such that it introduces a
convolution product on characters evaluating to the Taylor
coefficients $\delta_n^\phi,\delta_n^\psi$, such that the natural
group structure of these characters agrees with the diffeomorphism
group.

It turns out that this Hopf algebra of Connes and Moscovici is
intimately related to rooted trees in its own right \cite{CK1},
signalled by the fact that it is linear in generators on the rhs,
as are the coproducts of rooted trees and graphs.\footnote{Taking
the $\delta_n$ as naturally grown linear combination of rooted
trees imbeds the commutative part of the Connes-Moscovici Hopf
algebra in the Hopf algebra of rooted trees, which on the other
hand allows for extensions similar to the ones needed by Connes
and  Moscovici. Details are in \cite{CK1}.} This initiated the
collaboration of Alain Connes and the author, when, in a lucky
accident, we both stumbled over similar Hopf algebras at about the
same time.

Now, following \cite{CK3}, let us specialize to the massless case.
Then the formula for the bare coupling constant, \be g_0 = g \,
Z_1 \, Z_3^{-3/2} \ee (where both $g \, Z_1 =g + \d g$ and the
field strength renormalization constant $Z_3$ are thought of as
power series (in $g$) of elements of the Hopf algebra $H$) does
define a Hopf algebra homomorphism, $$ H_{CM} \build
\longra_{}^{g_0} H \, , $$ from the Hopf algebra $H_{CM}$ of
coordinates on the group of formal diffeomorphisms of $\Cb$ (ie
such that $ \vp (0) = 0 \, , \ \vp' (0) = {\rm id}$ as in
Eq.(\ref{expa})) to the Hopf algebra $H$ of the massless
theory.\footnote{We restrict ourselves to the massless theory so
that we can deal with one-dimensional diffeomorphisms. We can
regard a mass as a further coupling constant of a two-point vertex
which leads to formal diffeomorphisms of higher dimensional
spaces.} Having this Hopf algebra homomorphism from $H_{CM}$ to
$H$, dually one gets a transposed group homomorphism $\rho$, a
homomorphism from the huge group of characters of the Hopf algebra
to the group of diffeomorphism of physical parameters \cite{CK3}.
We finally recover the usual $\beta$-function: the image by $\rho$
of the previously introduced generator  $\b = Y \, {\rm Res} \,
\g$ is then the usual $\b$-function of the coupling constant $g$.
While this might sound rather abstract, it can be easily
translated into the standard notions of renormalization theory
(see, for example, \cite{Kra}).

While in \cite{CK3} the physical parameter under consideration was
a single coupling, similar considerations apply to other physical
parameters which run under the renormalization group, making use
of the Hopf algebraic description of composition of
diffeomorphisms in general.

As a corollary of the construction of $\rho$ one gets an {\it
action} by (formal) diffeomorphisms of the group $G$ on the space
$X$ of (dimensionless) coupling constants of the theory. One can
then in particular formulate the Birkhoff decomposition {\it
directly} in the group $ {\rm Diff} \, (X) $ of formal
diffeomorphisms of the space of coupling constants.

The unrenormalized theory delivers a loop $$ \d (\ve) \in {\rm
Diff} \, (X) \, , \ \ve \ne 0,  $$ whose value at $\ve\not=0$ is
simply the unrenormalized effective coupling constant. The
Birkhoff decomposition $ \d (\ve) = \, \d_+ (\ve) \, \d_-
(\ve)^{-1}  $ of this loop gives directly $$ \d_- (\ve) = \hbox{
bare coupling constant} $$ and $$ \d_+ (\ve) = \hbox{renormalized
effective coupling constant.} $$ This result is now stated in a
manner independent of our group $G$ or the Hopf algebra $H$, its
proof makes heavy use of these ingredients though.

Finally, the Birkhoff decomposition of a loop, $ \d (\ve) \in {\rm
Diff} \, (X)  $ admits a beautiful geometric interpretation
\cite{CK3}, described in Fig.(\ref{pic}).
\bookfig{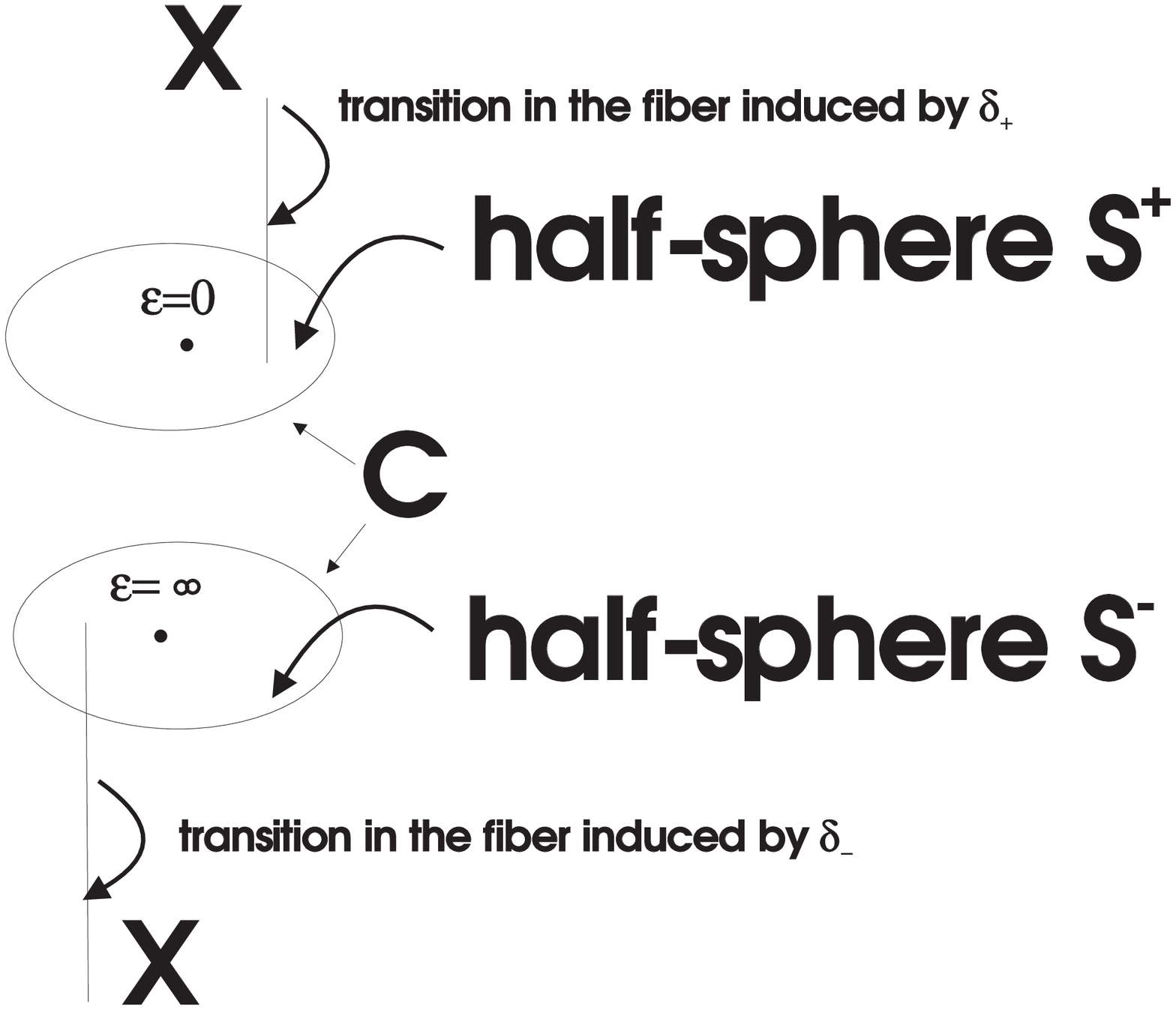}{}{pic}{The geometric picture of \cite{CK3}
allows for the construction of a complex bundle, $P = (S^+ \ts X)
\, \cup_{\d} (S^- \ts X)$ over the sphere $S = P_1 (\Cb) = S^+
\cup S^-$, and with fiber $X$, $ X \longra P \build
\longra_{}^{\pi} S \, ,$ where $X$ is a complex manifold of
physical parameters. The transition in this fiber are
diffeomorphisms. $\delta(\ve)$ delivers a diffeomorphism of $X$
for any $\ve\in C$, where $C$ is the boundary of the two
half-spheres $S^+,S^-$. It extends to the interiors of the
half-spheres via its Birkhoff decomposition. The meaning of this
Birkhoff decomposition, $ \d (\ve) = \d_+ \, (\ve) \, \d_-
(\ve)^{-1}$ is then exactly captured by an isomorphism of the
bundle $P$ with the trivial bundle, $ S \ts X.$ Note that
$\delta_-(\infty)$ is well-defined due to the fact that $S_{MS}$
has no constant term in $\ve$, which characterizes a minimal
subtraction scheme.}{8}
\subsection{An example}
In \cite{CK3} the reader can find explicit computational examples
up to the three-loop level, and a complete proof to all loop
orders, for the group and Hopf algebra homomorphisms described
above. We only want to check the Hopf algebra homomorphism $
H_{CM}\to H$ up to two loops here. We regard $g_0$ as a series in
a variable $x$ (which can be thought of as a physical coupling) up
to order $x^4$, making use of $g_0=xZ_1Z_3^{-3/2}$ and the
expression of the $Z$-factors in terms of 1PI Feynman graphs. The
challenge is then to confirm that the coordinates
$\delta_n^{g_0}$, implicitly defined by \cite{CM}
$$\log\left[g_0(x)^\prime\right]^{(n)},$$  as expected from
Eq.(\ref{CMT}), commute with the Hopf algebra homomorphism:
calculating the coproduct $\Delta_{CM}$ of $\delta_n$ and
expressing the result in terms of Feynman graphs with the help of
the character corresponding to $g_0$, $\tilde{
g}_0(\delta_n)=\delta_n^{g_0}$, must equal the application of the
coproduct $\Delta$ applied to $\delta_n^{g_0}$.

We write $g_0=xZ_1Z_3^{-3/2}$,
\[
Z_1=1+\sum_{k=1}^\infty z_{1,2k}x^{2k},
\]
\[
Z_3=1-\sum_{k=1}^\infty z_{3,2k}x^{2k},
\]
and
\[
Z_g=Z_1Z_3^{-3/2},\;z_{i,2k}\in {\cal H}_c,\;i=1,3,
\]
as formal series in $x^2$. Using
\[
\log\left(\frac{\partial}{\partial x}
xZ_g\right)=\sum_{k=1}^\infty\frac{\delta_{2k}^{g_0}}{(2k)!}x^{2k},
\]
which defines $\delta_{2k}^{g_0}$, we find
\begin{eqnarray}
\frac{1}{2!}\delta_2^{g_0} \equiv\tilde{\delta_2}^{g_0} & = &
3z_{1,2}+\frac{9}{2}z_{3,2},\label{e1}\\
\frac{1}{4!}\delta_4^{g_0} \equiv\tilde{\delta_4}^{g_0} & = &
5[z_{1,4}+\frac{3}{2}z_{3,4}]-\frac{9}{2}z_{1,2}^2
-6z_{1,2}z_{3,2}-\frac{3}{4}z_{3,2}^2,\label{e2}.\nonumber
\end{eqnarray}
The algebra homomorphism ${ H}_{CM}\to{ H}$ is effected by
expressing the $z_{i,2k}$ in Feynman graphs, with 1PI graphs with
three external legs contributing to $Z_1$, and 1PI graphs with two
external legs, self-energies, contributing to $Z_3$.

Explicitly, we have  {\large
\begin{eqnarray*} z_{1,2} & = & \v ,\\ z_{3,2}
&= & \frac{1}{2} \p ,\\ z_{1,4} &= &
\vlv+\vuv+\vdv+\frac{1}{2}\left[\vrp+\vup+\vdp\right]+\frac{1}{2}\w,\\
z_{3,4} &= & \frac{1}{2}\left[\pdp+\pv\right].\\
\end{eqnarray*}}
On the level of diffeomorphisms, we have the coproduct
\begin{eqnarray}
\Delta_{CM}[\delta_4] & = &
\delta_4\ot1+1\ot\delta_4+4\delta_2\ot\delta_2,\label{ed6}
\end{eqnarray}
where we skip odd gradings (in $\phi^3$ theory, adding a loop
order increases the order in the coupling by $g^2$).

We have to check that the coproduct $\Delta$ of Feynman graphs
reproduces these results.

Applying $\Delta$ to the rhs of (\ref{e2}) gives, using the
expressions for $z_{i,k}$ in terms of Feynman graphs,
 {\large\begin{eqnarray*}
\Delta(\tilde{\delta_4}) & = &
6\v\ot\v+\frac{9}{2}\left[\v\ot\p+\p\ot\v\right]\\
 & & +\frac{27}{8}\p\ot\p+\tilde{\delta_4}\ot 1+1\ot
\tilde{ \delta_4}.
\end{eqnarray*}}
This has to be compared with $\tilde{\delta_4}\ot 1+1\ot \tilde{
\delta_4}+\frac{2!2!}{4!}4\tilde{\delta_2}\ot\tilde{\delta_2}$,
which matches nicely, as
 {\large\begin{eqnarray*}
\tilde{\delta_2}\ot\tilde{\delta_2} & = & 9\v\ot\v+\frac{27}{4}
\left[\v\ot\p\;+\;\p\ot\v\right]\\ & & +\frac{81}{16}\p\ot\p.
\end{eqnarray*}}

\section{Conclusions and Outlook}
In this final section we mainly want to comment on some more
future lines of investigation, which in part are already work in
progress. We start with the connection between Feynman diagrams
and the numbers which we see in their coefficients of ultraviolet divergence,
which is a rich source of structure \cite{Book}.
\subsection{Numbers and Feynman diagrams}
There is an enormous amount of interesting number theory in
Feynman diagrams \cite{BKold1,BKold2,Book}. In particular, the
primitive elements in the Hopf algebra, those graphs which have no
subdivergences and provide a renormalization scheme independent
coefficient of ultraviolet divergence, show remarkable and hard to
explain patterns. These coefficients evaluate in Euler--Zagier
sums (generalized polylogs evaluated at (suitable roots of) unity
so that they  generalize multiple zeta values (MZVs)
\cite{Book,BKold1,BKold2}), numbers which have remarkably
fascinating algebraic structure \cite{Ghon,BBB,Don,MEH1}.

These algebraic structures are believed to be governed by shuffle
algebras, and by the much more elusive Grothendieck--Teichm\"uller
group (see, for example, \cite{dror1} for an introduction to the
Grothendieck Teichm\"uller group which is close in spirit to the
consideration of  short-distance singularities).

The coefficients of UV-divergence in Feynman diagrams typically
evaluate, up to the six loop level, in terms of these
Euler--Zagier sums, but the question if this will always be so
remains open in light of the failure to identify all these
coefficients in this number class at the seven loop level
\cite{BKold1,BKold2,Book}. While the embarrassingly successful
heuristic approach summarized in \cite{Book}, providing a
knot-to-number dictionary for those numbers, only emphasizes the
need for a more thorough understanding, the algebraic structures
in Feynman graphs hopefully lead to such an understanding in the
future. It is already remarkable that shuffle products can be
detected in Feynman graphs \cite{DK3}, but their are hints for
much more structure \cite{DK4}.

But while the existence of shuffle algebras in Feynman graphs can
essentially be straightforwardly addressed due to the fact that a
shuffle algebra makes use of the $B_+,B_-$ operators in a natural
way \cite{DK3}, these remaining algebraic relations between
Feynman graphs will be harder to address.\footnote{But note that
these shuffle algebras and shuffle identities only hold for the
coefficients of ultraviolet singularity: they hold up to finite
parts, up to finite rerenormalizations that is.} But the very fact
that Feynman graphs realize their short-distance singularities in
tree-like hierarchies suggests that they can be understood along
lines similar to what is known for Euler--Zagier sums.

In particular, Feynman graphs whose subdivergences realize the
same rooted tree but with subgraphs inserted at different internal
lines provide remarkable number-theoretic features \cite{DKold}.
As mentioned before, in the operad picture, such differences are
given by permutations $\sigma(i)=j$ of places $i$ at which we
compose: $$\Gamma\circ_i\gamma\to\Gamma\circ_{\sigma(i)}\gamma.$$
Note that, if we let $U$ be the difference of the two expressions,
we get a primitive element in the Hopf algebra (if the two graphs
$\Gamma$ and $\gamma$ are both primitive), $\Delta(U)=U\otimes
1+1\otimes U$.

Quite often, one finds that these differences are even finite,
which means that the coefficients of ultraviolet divergence are
the same and drop out in the difference: short distance
singularities are invariant under the above permutations.
Fig.(\ref{galois}) gives an example of such an invariance observed
in \cite{DKold}. We insert a one-loop bubble at different places
$i,j$ in the graph. We do not have to worry that in one case it is
a one-loop fermion self-energy, in the other case a one-loop boson
self-energy. In massless Yukawa theory, they both evaluate to the
same analytic expression. This makes it very easy to study the
effect of a subdivergence being inserted at different places in a
larger graph. In this four-loop example, the difference becomes a
primitive element and hence delivers only a first order pole
$\sim\zeta(3)/\ve$, signalling the difference in topology between
the two diagrams \cite{Book}. The ladder diagram evaluates to
rational coefficients in the poleterms of its MS counterterm,
while the other diagram has the same rational part, but also has
$\zeta(3)$ in the $1/\ve$ pole. In the difference, only this first
order pole $\sim \zeta(3)/\ve$ remains.

Comparing the two three-loop subgraphs of each diagram, one finds
their difference to be finite and $\sim\zeta(3)$, so that the
three coefficients $\sum_{i=1}^3 c_i/\ve^i$ are invariant under an
exchange of the place where we insert the subgraph: the morphism
sending one graph to the other, and thus sending one configuration
of internal vertices with its characteristic short-distance
singularities to another, is a finite one. Similar observation
hold for higher loop orders \cite{DKold}.

A systematic understanding of such phenomena, and a possible
relation to finite-type invariants, seems crucial to understand
the algebraic relations in Feynman graphs completely. Ultimately,
one hopes for a geometric understanding of the analytic challenge
posed by Feynman diagrams. Meanwhile, similar relations have been
observed in QED \cite{DiplB}. \bookfig{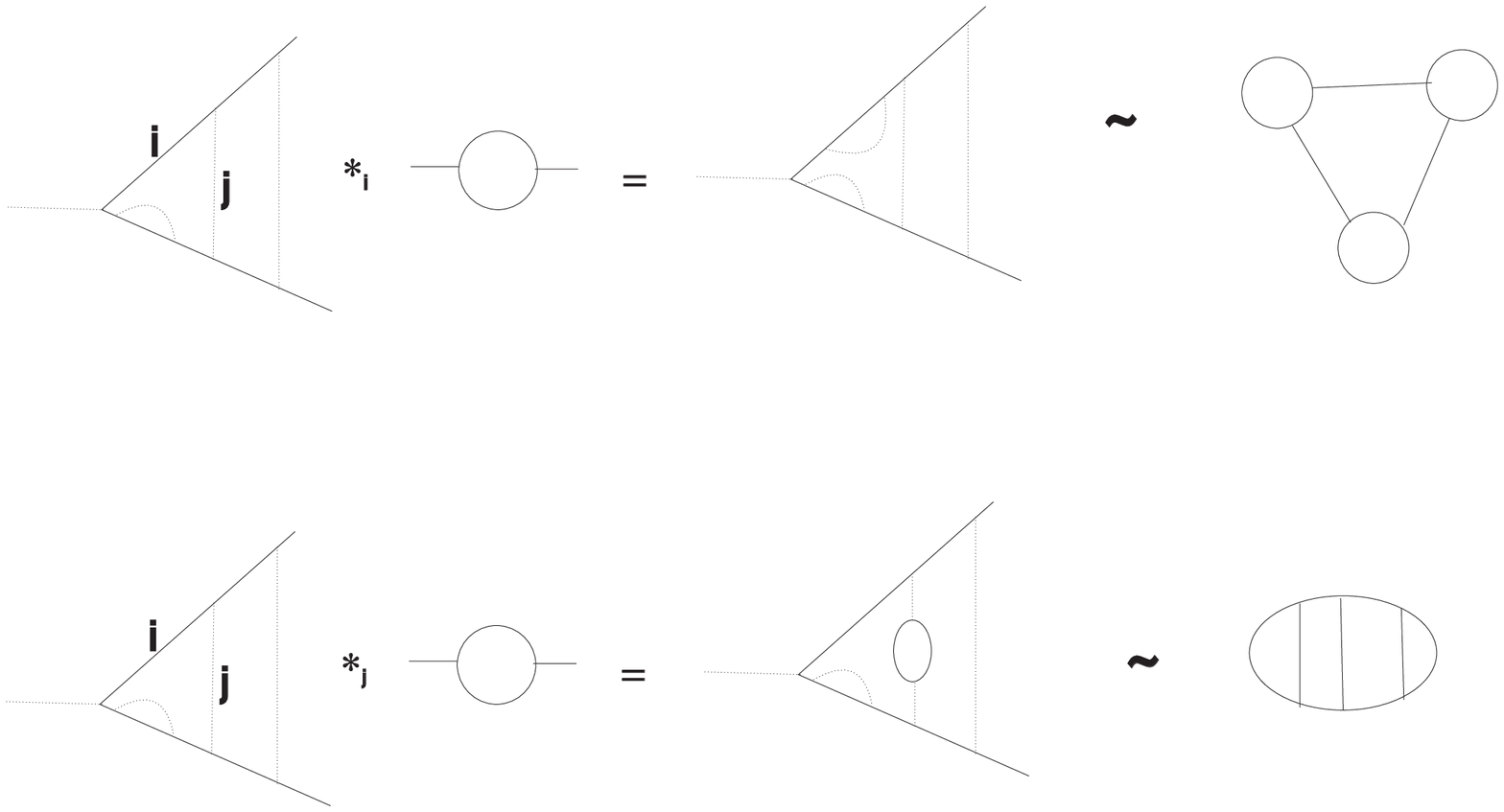}{}{galois}{These
two Feynman graphs (with their distinct topologies indicated on
the rhs of each: the topology of the upper graph is that of
disjoint one-loop insertions, the lower is a ladder topology) in
massless Yukawa theory have a remarkable relation: their
difference is a primitive Hopf algebra element. When evaluating
the character $S_{MS}$ on both, one finds a Laurent series with
poles of fourth order from both of them. In the difference, the
highest three-pole terms drop out, and the remaining term is
$\sim\zeta(3)/\ve$. Similar phenomena happen at higher loop orders
\cite{DKold}: higher poleterms are invariants under the
permutation of places where we insert subgraphs.}{6}

A requirement on the way to such an understanding is the question
how in the geometric picture of Fig.(\ref{pic}) one can relate an
infinitesimal variation in the base space to a variation in the
fiber, ie the quest for a connection?

For the $\delta_-$ part of the Birkhoff decomposition, this leads
to an investigation as to how a derivative with respect to the
regularization parameter $\ve$ is related to the insertion of a
further graph. First results at low loop orders to be discussed
elsewhere indicate that this is a source for relations between the
coefficients of ultraviolet divergence similar but not quite like
the four-term relations discussed in the study of finite-type
invariants \cite{Book}. This is not impossible: while all higher
poleterms are fixed in terms of the residue by the scattering type
formula Eq.(\ref{scatt}) of the previous section, this formula can
by its very nature not deliver relations between residues of
graphs.
\subsection{Gauge symmetries}
Clearly, one of the most urgent and fascinating questions is the
role of symmetries in quantum field theories. Having, with the
Hopf algebra structures reported here, discovered such a wonderful
machinery which encapsulates the quest for locality, one should
expect interesting structure when considering local gauge
symmetries, in particular also with respect to the role which
foliations play naturally in noncommutative geometry
\cite{CM,survey}.

There are many aspects which can hopefully be addressed in the
near future. \begin{itemize}
\item To what extent can Ward- and
Slavnov-Taylor identities be incorporated in this picture? Do
these identities form something like an ideal in the algebra of
graphs? Note that the language of external structures allows
nicely to formulate concepts like the longitudinal and transversal
part of a vertex-correction for example, and is hence well-adopted
to address such questions.

\item Has BRST cohomology a natural formulation in this context?
\item Gauge theories provide number-theoretical miracles in abundance,
with the most significant observation being Jon Rosner's
observation \cite{Rosner} of the vanishing of $\zeta(3)$ from the
$\beta$-function of quenched QED. While this can be understood
heuristically \cite{BDK,Book},
eventually the role between internal symmetries and
number-theoretic properties must be properly understood.
\end{itemize}
For the practitioner of quantum field theory, the real challenge
lies in the treatment of the perturbative expansion in
circumstances when there is no regularization available which
preserves the symmetries of the initial theory. A notorious and
famous problem at hand is the $\gamma_5$ problem in dimensional
regularization \cite{Fred}. In realistic circumstances like the
Standard Model this already demands a formidable effort at the
one-loop level if one uses a calculational scheme which violates
the BRS symmetry even in the absence of anomalies (see
\cite{Martin} for such an example), which then is an unavoidable
effort dictated by the demand to restore the BRS symmetry using
the quantum action principle. There is one obvious useful role for
the Hopf algebra: the analysis at the one-loop level would in many
ways not change when extended to any other primitive element of
the Hopf algebra, which, being primitive,  all share with the
one-loop graphs that they have no subdivergences. From there, the
Hopf algebra structure governs the iteration of graphs into each
other.

But then, the prominent role and natural role which
field-theoretic ingredients like the Dirac propagator and
$\gamma_5$ itself, a volume form on four-dimensional space
essentially, play in non-commutative geometry \cite{CKl,survey},
gives hope for a more profound understanding of this problem in
the future.

\subsection{The exact renormalization group and the non-pertur\-ba\-tive regime}
Ultimately, the renormalization group is a non-perturbative
object, and can indeed be addressed without necessarily making use
of the usual concepts of graph-theoretic expansions
\cite{Polch,BB}. This is nicely reflected by the fact that the
transition from the perturbative to the non-perturbative just
amounts, in the picture outlined here, to a Birkhoff decomposition
of an actual instead of a formal diffeomorphism. Integrationg out high frequency modes
in the functional integral step-by-step produces a sequence of
diffeomorphisms of the correlation function under
consideration.\footnote{The fact, emphasized by Polchinski \cite{Polch},
 that in such an approach one does not see the graph-theoretical
notions emphasized in textbook approaches to renormalization
theory is a mere reflection of the fact that one can formulate the
Birkhoff decomposition directly on the level of diffeomorphisms of
physical observables \cite{CK3}, as exhibited in the previous
section.}

The Hopf algebra of rooted trees, thanks to its universality,
provides the relevant backbone in any case, and indeed rooted
trees underly any iterative equation, like, for example, the
Wilson equation $$\frac{\partial S_\lambda}{\partial
\lambda}={\cal F}(S_\lambda),$$ for some action parametrized by
some cut-off $\lambda$ and some suitable functional ${\cal F}$.
Integrating this functional ${\cal F}$ now plays the role of the
operator $B_+$ in the universal setting of the Hopf algebra of
rooted trees \cite{CK1}. Rooted trees are deeply built into
solutions of (integro-) differential equations
\cite{Butcher,Brou}. It is no miracle then that on the other hand
one finds that the understanding of the Hopf- and Lie algebras of
Feynman graphs not only enables high-loop order calculations
\cite{BK1,BK2,BK4} which allow to analyze Pad\'e-Borel
resummations \cite{BK2,BK4,JS} but also allows to find exact
non-perturbative solutions in new problems. A first result can be
found in \cite{BK4}.

\subsection{Further aspects}
Combinatorially, rooted trees are very fundamental objects, and
their Hopf and Lie algebra structure underlies not only the
combinatorial process of renormalization, but can hopefully be
used in the future in other expansions in perturbation theory,
starting from a disentanglement of infrared divergent sectors
\cite{BH} to more general applications in asymptotic expansions
\cite{smirnov}. Its universal nature already allowed to use it in
a straightforward  formulation of block spin transformations,
coarse graining and the renormalization of spin networks
\cite{Fot}. Eventually, one hopes, this basic universal
combinatorial structure finds its way into other approaches to
QFT, from the constructive approach \cite{Jaffe} which in its
nature is very tree-like from a start \cite{Riv}, to the algebraic
school \cite{BV,DF}, which all have to handle the basic
combinatorial step that we can address a problem only after we
addressed its subproblems.\footnote{The universality of the Hopf
algebra can be used to describe effective actions in a unifying
manner, which was indeed one of the main points of \cite{CK2,CK3},
while the connection to integrable models promoted in
\cite{GMS,MM} can hopefully  be substantiated further in the
future.} Note also that applications of forest formulas in the
context of noncommutative field theory and string field theory
(see \cite{Chep} for a detailed graphical analysis) naturally
change the criteria for the subgraphs $\gamma$ over which a sum
$$\Delta(\Gamma)=\Gamma\otimes
1+1\otimes\Gamma+\sum_{\gamma}\gamma\otimes\Gamma/\gamma$$ runs,
while the results in \cite{DKo} underline that a Hopf algebra
structure can still be established when we vary these criteria.

There is no space here to comment in detail on some other
mathematical developments which are related to the discovery of
the Hopf algebra structure of renormalization. We can only address
the interested reader to \cite{Moe,Pan,Reims,CL}. But note that
such mathematical investigations are often very useful for a
practitioner of QFT: clearly, the classification of all primitive
Hopf algebra elements is of importance even for the case of the
undecorated Hopf algebra of rooted trees, and leads for example to
the notion of a bigrading which characterizes potential higher
divergences algebraically \cite{BK3,Reims}.

\subsection{Conclusions}
Rooted trees and Feynman graphs are familiar objects for anybody working
on the perturbative expansion of a functional integral, and as familiar are
forest formulas and the Bogoliubov recursion.

What is new is that there is a universal Hopf algebra on rooted
trees, devoted to the problem of singularities along diagonals in
configuration spaces and providing a principle of multiplicative
subtraction, which reproduces just these recursions and forest
formulas. That Feynman graphs, with all their external structure,
form a Lie algebra is a very nice consequence which hopefully
gives a new and strong handle for the understanding of QFT in the
future. The consequences of the connection to the Riemann--Hilbert
problem and the Birkhoff decomposition of diffeomorphisms, the
connection between short-distance singularities in perturbation
theory and polylogarithms, all this indicates what a rich source
of mathematical structure and beauty is imposed on a quantum field
theory by its infinities.

\section*{Acknowledgments}
A large body of the work presented here was done in past and
ongoing collaborations with David Broadhurst and Alain Connes.
Helpful discussions with Jim Stasheff on operads are gratefully
acknowledged. This work was done in part for the Clay Mathematics
Institute. Also, the author thanks the DFG for a Heisenberg
fellowship.

\end{document}